\input epsf.sty
\documentstyle[prd,aps]{revtex}

\firstfigfalse
\newcommand{\incpsf}[1]{
\centerline{\epsfysize=6.0in \epsfxsize=6.0in \epsfbox{#1}}}
\newcommand{\incpsfw}[1]{\centerline{\epsfxsize=6.0in \epsfbox{#1}}}

\begin{document}

\draft

\title{Numerical evolution of 
matter in dynamical axisymmetric black hole spacetimes.
I. Methods and tests}

\author{S. Brandt${}^{1}$,
 J.A. Font${}^{1}$,
 J.M$^{\underline{\mbox{a}}}$ Ib\'a\~nez${}^{2}$,
 J. Mass\'o${}^{1,3}$ and E. Seidel${}^{1,4}$}

\address{${}^{1}$ Albert Einstein Institute \\
Max-Planck-Institut f\"ur Gravitationsphysik \\
Schlaatzweg 1, 14473 Potsdam, Germany}
\address{${}^{2}$ Departamento de Astronom\'{\i}a y Astrof\'{\i}sica \\
Universidad de Valencia, 46100 Burjassot (Valencia), Spain}
\address{${}^3$ Departament de F\'\i sica, Universitat de les Illes
Balears, 07071 Palma de Mallorca, Spain}
\address{${}^4$ Departments of Astronomy and Physics\\
and\\
National Center for Supercomputing Applications,\\
Champaign, IL, 61820, USA}

\date{\today}

\maketitle
\begin{abstract}

We have developed a numerical code to study the evolution of 
self-gravitating matter in dynamic black hole axisymmetric spacetimes 
in general relativity.  The matter fields are evolved with a 
high-resolution shock-capturing scheme that uses the characteristic 
information of the general relativistic hydrodynamic equations to 
build up a linearized Riemann solver.  The spacetime is evolved with 
an axisymmetric ADM code designed to evolve a wormhole in full general 
relativity.  We discuss the numerical and algorithmic issues related 
to the effective coupling of the hydrodynamical and spacetime pieces 
of the code, as well as the numerical methods and gauge conditions we 
use to evolve such spacetimes.  The code has been put through a series 
of tests that verify that it functions correctly.
Particularly, we develop and describe a new set of testbed calculations and 
techniques designed to handle dynamically sliced, 
self-gravitating matter flows on black holes, and subject the code 
to these tests.  We make some studies of the spherical and axisymmetric 
accretion onto a dynamic black hole, the fully dynamical evolution of 
imploding shells of dust with a black hole, the evolution of matter 
in rotating spacetimes, the gravitational radiation induced by the 
presence of the matter fields and the behavior of apparent horizons 
through the evolution.

\end{abstract}

\pacs{PACS numbers: 95.30.Sf, 95.30.Lz, 97.60.Lf,
04.25.Dm, 04.40.-b, 04.30.Db, 47.11.+j, 47.75.+f}

\section{Introduction}
\label{intro}

\subsection{Overview}

An accurate description of the evolution of matter in a fully 
dynamical spacetime in complete generality is a longstanding and 
still unresolved problem in numerical relativistic astrophysics.  With 
many important problems in urgent need of 
study, such as multidimensional 
core collapse, neutron star collisions or black hole 
formation, this outstanding problem represents a major gap in our 
ability to understand some of the most highly energetic 
processes in relativistic
astrophysics. Numerous attempts have been made to develop such 
capabilities over the years, but only the spherical case can be 
considered essentially 
solved~\cite{May66,Dykema80,Shapiro80,Evans84,Sch88,Mezz89,Gou91,Romero96},
and even there it has not yet been widely applied.  In 
higher dimensions, most of the studies have been restricted to the 
axisymmetric two dimensional (2D) case, and there much of the work has 
been devoted to hydrodynamical integrations
on {\em fixed} general relativistic 
backgrounds\cite{Hawley84,Hawley84a,Petrich89,Banyuls97,Font98a}.  
Even in axisymmetry, 
only a few attempts have been made to consider a fully 
self-gravitating, dynamic 
background~\cite{Nakamura81,Stark85,Evans86,Abrahams94a}.  The 
three-dimensional case is also being studied, but the number of fully 
relativistic simulations is even smaller, with only a handful of fixed 
background hydrodynamical codes~\cite{Hawley91,Duncan93a} 
and fully self-gravitating codes~\cite{Oohara96} developed 
to date. In view of the difficulty of a fully 
relativistic treatment, approximations to a complete 
general relativistic approach are also being developed 
recently~\cite{Wilson95,Wilson96}.

The reasons for the reduced number of multidimensional (2D and 3D) 
computations are diverse, but we can enumerate the two most obvious 
ones: first, the inherent difficulties and complexities of the system 
of equations to integrate, the Einstein field equations coupled to the 
general relativistic hydrodynamic equations, which make this 
computation one of the most challenging ones in physics.  Second, the 
immense computational resources needed to integrate the equations for
3D evolutions were not available in the past and are only starting
to be achieved at present.  

\subsection{Axisymmetric General Relativistic Hydrodynamics}

In the axisymmetric case there exist a number of interesting 
astrophysical applications that can be addressed numerically, such as 
the rotational collapse of stellar cores in the supernova explosion 
scenario, the implosion of matter shells onto a black hole, the 
dynamics and stability of accretion disks or the fully dynamical 
spacetime version of the so-called relativistic Bondi-Hoyle accretion 
onto a moving black hole. Some of these problems have already
tentatively been studied in the past within the framework of general
relativity (see references below) although many
scenarios still await for first (and detailed) computations. In this paper
we take some early steps on that direction 
focusing on the methodology we will be using in the future for the study
of those systems.

In addition to the aforementioned scenarios, one of the most 
interesting systems to consider is the head-on collision of a 
neutron star with a black hole.  The correct understanding of this 
simplified head-on model will undoubtedly serve as an important testbed 
for future three-dimensional codes.  Hence, it is an important step 
that will aid in development of simulations of more complex scenarios, 
such as the coalescence and merging of spiral compact binaries.  These 
catastrophic events are believed to be among the most promising 
sources of gravitational radiation to be detected by the gravitational 
wave interferometers to be operating around the turn of the century 
(LIGO, VIRGO, TAMA, GEO600).

The use of general relativistic axisymmetric codes in numerical 
relativity has been largely devoted to the study of the gravitational 
collapse and bounce of a rotating stellar core and the subsequent 
emission of gravitational radiation.  These investigations started 
with the work of Nakamura~~\cite{Nakamura81}, who was the first to 
calculate a general relativistic rotating stellar collapse.  In his 
calculation, he was able to track the evolution of matter and the 
formation of a black hole but the scheme was not accurate enough to 
compute the emitted gravitational radiation.  Later, Stark and 
Piran~\cite{Stark85} revisited this problem studying the collapse of 
rotating relativistic polytropic stars to black holes and succeeded in 
computing the gravitational radiation emission.  Their code used the 
radial gauge and a mixture of polar and maximal slicing.  The 
hydrodynamic equations were solved with standard finite difference 
methods with artificial viscosity~\cite{Wilson72,Wilson79}.  
Evans~\cite{Evans86} also studied the gravitational collapse problem 
but for non-rotating matter configurations.  His numerical 
scheme to treat the matter fields was more sophisticated than previous 
approaches as it included monotonic upwind reconstruction procedures 
and flux limiters.  Evans~\cite{Evans86} was able to show that 
Newtonian gravity and the quadrupole formula for gravitational 
radiation were inadequate to study the problem.  
More recently, Abrahams et al.~\cite{Abrahams94a} have solved the Einstein 
equations numerically for rotating spacetimes where the source of the 
gravitational field is a configuration of collisionless (dust) 
particles.  They used the code to evaluate the stability of polytropic 
and toroidal star clusters. Because they did not have to take 
pressure forces into account, they could reduce the 
hydrodynamic computation 
to a straightforward integration of the geodesic equations.

On the other hand, a few groups have concentrated on building codes 
to handle dynamical, axisymmetric gravitational fields in vacuum, 
without the complications of the matter fields. 
These have been applied notably 
to the collapse of gravitational waves to form a black 
hole~\cite{Abrahams92b,Abrahams93a} and to the evolution of distorted, 
rotating, and colliding black 
holes~\cite{Bernstein93b,Bernstein94a,Anninos93b,Anninos94b}.
But, even without the presence of matter fields,
these calculations have proven very difficult 
for a number of reasons, not the least of which are the choice of 
appropriate gauge conditions and the presence of 
numerical instabilities encountered near the axis of symmetry.  These 
are among the reasons that work in this area has been fairly limited.

\subsection{Our proposal}

We have embarked on a program to develop a series of fully
relativistic codes to solve the coupled set of
equations in multiple dimensions.  This work builds on some years of
experience in dynamic vacuum relativity (see, for
example~\cite{Brandt94b,Brandt94c,Anninos94c,Anninos94d,Bona98b}) and
in fixed background general relativistic hydrodynamics
(see~\cite{Marti91,Banyuls97,Font98a,Font98c}). 
In spherical symmetry, we have
developed a code\cite{Font97,Aloy98a} exploiting advances in hyperbolic
treatments of both general relativistic 
hydrodynamics~\cite{Banyuls97} and
the Einstein equations themselves~\cite{Bona94b,Bona97a}.  In the 3D
case, a code is under development specifically for the study of
neutron star collisions~\cite{Bona98b,Font98b,Bougleux98}.            

In the present paper we report first results obtained with a 
2D axisymmetric numerical code designed to evolve 
rotating black hole spacetimes with self-gravitating matter.
We believe that 
this code can have a number of interesting applications in the
numerical modeling of accretion processes in black hole astrophysics.
In addition, it will provide 
important testbeds for fully 3D codes now under development.
This axisymmetric code is the result of 
the {\it coupling} of two previously developed production codes: an advanced 
general relativistic hydrodynamical code for stationary spacetimes and 
an axisymmetric black hole code to solve the 
dynamic Einstein field equations in vacuum. 
The integration of the different variables, spacetime and hydrodynamical
quantities, is, hence, performed with a unique code.
The two basic building blocks
of the new code are extensively described in previous 
work~\cite{Banyuls97,Brandt94a,Brandt94b,Brandt94c}. 

The equations of general relativistic hydrodynamics are integrated 
with a modern {\it high-resolution shock-capturing} numerical scheme 
which relies on the knowledge of the characteristic information of the 
system of equations in order to build up a {\it Riemann solver}.  
These equations have been recently written as a hyperbolic system of 
conservation laws (with sources) for a general $\{3+1\}$ 
metric~\cite{Banyuls97}.  The characteristic information of the system 
has been explicitly obtained.  The Jacobian matrices are found to have 
real eigenvalues and a complete set of right-eigenvectors, therefore 
satisfying the definition of hyperbolicity~\cite{Leveque92}.  The use 
of these methods, also referred to as Godunov-type methods (after the 
seminal idea by Godunov~\cite{Godunov59})
is becoming more important in recent years, 
due to a number of nice properties that other finite difference 
methods do not share (e.g., artificial viscosity methods).  Those 
include a consistent treatment of discontinuous solutions 
(shock-capturing property) combined with being high-order methods in 
regions where the numerical solution is smooth. Recently, the 
use of these methods in relativistic hydrodynamics has permitted, for 
the first time, an accurate description of ultrarelativistic flows in 
different astrophysical scenarios (see, e.g.,~\cite{Font98a},\cite{Marti97} 
and references therein).

The black hole piece of the code was originally developed by Brandt 
and Seidel~\cite{Brandt94b,Brandt94c} and is based on the standard 
Arnowitt-Deser-Misner (ADM)~\cite{Arnowitt62} formulation of the 
Einstein equations as an initial value (or Cauchy) problem.  The 
metric and extrinsic curvature components are evolved for the full set 
of the Einstein equations using a $\{3+1\}$ explicit second 
order Runge-Kutta scheme 
with centered differencing.  This code has a black hole built into the 
initial hypersurface of the spacetime.  This avoids possible 
coordinate problems at the origin of the spherical coordinate system, 
since the black hole is constructed with an {\em isometry} that maps 
its interior, which contains a singularity at the 
origin, to the exterior, across a sphere at a finite radius. Hence, no 
reference need be made to the origin (as discussed in more detail 
below). This code can evolve a variety of spacetimes including 
rotating, distorted black holes.  It also has a number of utilities 
built-in, such as routines to extract the various gravitational 
radiation modes and to track the motion of apparent and event 
horizons.

The effective coupling of the two systems is through the source terms 
of the Einstein field equations.  This allows us to integrate
the whole system in a straightforward way -- the metric and matter
codes can simply take turns updating the variables.
First, the hydrodynamical code 
takes a step, treating the spacetime metric as fixed and then the 
black hole code takes a step treating the matter fields as fixed.
We regard this coupling of two mature codes as 
a starting point for algorithm development and forthcoming work in 
numerical relativistic astrophysics that will be explored in the near
future.

The organization of the paper is as follows: In 
section~\ref{preliminaries} we briefly review the ADM formulation of 
the spacetime, discuss our particular choice of spacetime and 
hydrodynamical variables and describe the equations of general 
relativistic hydrodynamics.  The initial value (Cauchy) problem with 
the presence of matter fields is discussed in section~\ref{initdat}.  We 
describe the details of the numerical code in section~\ref{code}, where
we address the specific issues concerning the coupling of the 
hydrodynamic equations and Einstein equations, as well as the gauge 
conditions used.  Results and convergence tests of the simulations are 
presented in section~\ref{results}.  These include the spherical 
(Bondi) accretion of matter and the implosion (accretion)
of a dust shell with the 
black hole.  In the latter case, we also extract the waveforms of the 
gravitational waves induced by the presence of matter.  Although in 
the present paper we mainly focus on the non-rotating case we will 
also present some results for rotating black hole 
spacetimes at the end of section~\ref{results}.  Finally, 
section~\ref{conclusions} summarizes our main conclusions and outlines 
future directions of this work.

\section{Preliminaries}
\label{preliminaries}
\subsection{ADM formulation of spacetime}

We use the standard ADM~\cite{Arnowitt62} formulation of the Einstein
equations as the basis for our numerical code.  Pertinent details of
the formalism are summarized here, but we refer the reader to
Ref.~\cite{York79} for a general treatment.  A separate description of
the theoretical details of both parts of the code can be found in
Refs.~\cite{Brandt94b} and~\cite{Banyuls97}.  Although 
for the sake of completeness we outline the most
important points here, we refer the interested reader to these
references as they will be extensively used throughout the present
paper.

In the ADM formulation, spacetime is foliated into a set of
non-intersecting spacelike hypersurfaces.  There are two kinematic
variables which describe the evolution between these surfaces: the
lapse function $\alpha$, which describes the rate of advance of time
along a timelike unit vector $n^\mu$ normal to a surface, and the
spacelike shift vector $\beta^i$ that describes the motion of
coordinates within a surface (throughout the paper Greek (Latin)
indices run from $0$ to $3$ ($1$ to $3$)).  The choice of the lapse
function and shift vector is essentially arbitrary (i.e., four degrees
of freedom embodying the coordinate freedom of general relativity),
and our choices will be described in section~\ref{gauge} below.

The line element is written as
\begin{equation}
 ds^2 = -(\alpha^{2} -\beta _{i}\beta ^{i}) dt^2 + 2 \beta _{i} dx^{i}
 dt +\gamma_{ij} dx^{i} dx^{j},
\end{equation}
where $\gamma_{ij}$ is the 3--metric induced on each spacelike slice.
Given a choice of lapse $\alpha$ and shift vector $\beta^{i}$, the
Einstein equations in the $\{3+1\}$ formalism split into evolution
equations for the 3--metric $\gamma_{ij}$ and constraint equations
that must be satisfied on any time slice. The evolution equations are
\begin{eqnarray}
\partial_t \gamma_{ij} & = &
-2 \alpha K_{ij}+\nabla_i \beta_j+\nabla_j \beta_i \\
\partial_t K_{ij} &=&
-\nabla_i \nabla_j \alpha + \alpha \left( R_{ij}+K\ K_{ij} -2 K_{im}
K^m_j \right) \nonumber \\ &+& \beta^m \nabla_m K_{ij} + K_{im}
\nabla_j \beta^m+K_{mj}
\nabla_i \beta^m \nonumber \\
&-&8 \pi \alpha \left(T_{ij}-\frac{1}{2}\gamma_{ij} T_m^m+\frac{1}{2} 
\rho_E \gamma_{ij}\right)
\label{excurv}
\end{eqnarray}
where $K_{ij}$ is the extrinsic curvature of the 3--dimensional time slice,
$R_{ij}$ is the Ricci tensor of the induced 3--metric and
$\nabla_i$ is the covariant three-space derivative.  
The matter terms involving $T_{ij}$ and $\rho_E$ are defined below.

The Hamiltonian constraint equation is
\begin{equation}
R+K^2-K^{ij} K_{ij}=16 \pi \rho_E\label{hamconstrnt}
\end{equation}
with $R$ being the Ricci scalar.  The three momentum constraint
equations are
\begin{equation}
\nabla_i \left(K^{ij}-\gamma^{ij} K \right)=8 \pi S^j.\label{momconstrnt}
\end{equation}

In the above equations, the quantities $T^{ij}$, $S^i$ and $\rho_E$, the
spatial components of the stress-energy tensor, the momenta and the
total energy, respectively, are obtained by projecting the four
stress-energy tensor using $n_\mu$, the normal to the slice:
\begin{eqnarray}
\perp^i{}_\mu &=& g^i{}_\mu + n^i n_\mu \\
T^{ij} &=& \perp^i{}_\mu \perp^j{}_\nu {}^{4}T^{\mu \nu}  \\
S^i &=& - \perp^i{}_\mu {}^{4}T^{\mu \nu} n_{\nu} \\
\rho_E &=& \,\,\,{}^{4}T^{\mu \nu} n_\mu n_\nu.
\label{defTSE}
\end{eqnarray}

Note that Latin indices are raised and lowered with the
induced 3--metric, e.g., $K=K_{ij} \gamma^{ij}$, $\beta_i =
\beta^j \gamma_{ij}$. As we discuss in section~\ref{initdat}, the
constraint equations are used to obtain the initial data, and the
evolution equations are used to advance the solution in time.

\subsection{Hydrodynamic Equations and Variables}
\label{hydrodynamics}

The matter fields appearing in the constraint and extrinsic curvature
equations are computed via the local conservation laws of baryon
number and energy-momentum,
\begin{equation}
\nabla_{\mu} J^{\mu} = 0
\label{J}
\end{equation}
\noindent
\begin{equation}
\nabla_{\mu} T^{\mu \nu} = 0.
\label{T}
\end{equation}

The rest-mass current and the 
perfect fluid stress-energy tensor of Eqs.~(\ref{J}) and~(\ref{T}) have the
following definitions
\begin{equation}
J^{\mu} = \rho u^{\mu}
\end{equation}
and
\begin{equation}
T^{\mu\nu} = \rho h u^{\mu}u^{\nu} + p g^{\mu \nu}\label{TrhouP}.
\end{equation}
\noindent
Here, $\rho$ is the rest-mass density, $p$ is the pressure,
$u^{\mu}$ is the four-velocity of the fluid and $h$ is
the specific enthalpy, defined by
$h = 1 + \varepsilon + p/\rho$, where $\varepsilon$ is 
the specific internal energy. The spatial part of the fluid velocity is
defined according to
\begin{equation}
u^\mu = W \left(v^\mu+n^\mu\right)\label{3vel}
\end{equation}
such that $v^\mu n_\mu=0$.  Given this condition, it follows that
$W$ is the Lorentz factor, $W = -u^\mu n_\mu = 1/\sqrt{1-v_\mu v^\mu}$
and
\begin{eqnarray}
\perp^i{}_\mu u^\mu &=& W v^i, \\
\perp^i{}_\mu \perp^{}_\nu u^\mu u^\nu &=& W^2 v^i v^j.
\end{eqnarray}
In Ref.~\cite{Banyuls97} the general relativistic hydrodynamical equations
are written, explicitly, as a hyperbolic system of balance laws
in the framework of the ADM formulation.
Starting from
Eqs.~(\ref{J}) and (\ref{T}) and choosing an appropriate basis vectors
for the spacetime $e^{(i)}_\mu$,
adapted to the Eulerian observers it is possible to cast the
conservation equations into a more useful form~\cite{Banyuls97}:
\begin{equation}
\nabla_\mu
\left[ J^\mu,
T^{\mu\nu} e^{(i)}_\nu,
T^{\mu\nu} n_\mu-J^\nu\right]
=
\left[0,
T^{\mu\nu} \nabla_\mu e^{(i)}_\nu,
T^{\mu\nu} \nabla_\mu n_\nu \right]
\end{equation}
We can adopt a more compact notation as follows:
\begin{eqnarray}
\nabla_\mu {\bf f}^\mu &=& {\bf s} \\
{\bf f}^\mu &=&
\left[ J^\mu,
T^{\mu\nu} e^{(i)}_\nu,
T^{\mu\nu} n_\mu-J^\nu\right]\\
{\bf s} &=& \left[0,
T^{\mu\nu} \nabla_\mu e^{(i)}_\nu,
T^{\mu\nu} \nabla_\mu n_\nu \right].
\end{eqnarray}
\noindent
Finally, this may be written in flux form by defining
\begin{equation}
{\bf u} = -{\bf f}^\mu n_\mu/\alpha
\end{equation}
giving us
\begin{equation}
\frac{1}{\sqrt{-g}} \left(
\frac {\partial \sqrt{\gamma}{\bf u}}
{\partial x^{0}} +
\frac {\partial \sqrt{-g}{\bf f}^{i}}
{\partial x^{i}} \right)
 = {\bf s}.
\label{F}
\end{equation}
In the above equation $\gamma$ stands for the determinant of the 3-metric
and $\sqrt{-g}=\alpha\sqrt{\gamma}$.
Thus, the matter variables that are actually evolved in time are
\begin{eqnarray}
{\bf u} &=& \left[D,S^i,\tau\right] \nonumber\\
&=& \left[
J^\mu n_\mu,
-\perp^i{}_\nu T^{\nu\mu} n_\mu,
T^{\mu\nu} n_\mu n_\nu-J^\mu n_\mu \right] \nonumber\\
&=& \left[ \rho W, \rho h W^2 v^i,\rho h W^2-\rho W-P \right].
\end{eqnarray}
\noindent
For numerical applications we give the equations the form of a {\em canonical}
balance law.  Hence, we move quantities related to the spacetime
metric to the RHS of Eq.~(\ref{F}). Explicitly, we have
\begin{eqnarray}
\frac{\partial {\bf u}}{\partial t} +
\frac{\partial (\alpha {\bf f}^i)} {\partial x^i} =
{\tilde {\bf s}}
\label{hydrosystem}
\end{eqnarray}
\noindent
with
\begin{eqnarray}
{\tilde {\bf s}} =
\alpha {\bf s} -
\frac{ {\bf u}}{\sqrt{\gamma}} \frac{\partial\sqrt{\gamma}}
{\partial t} -
\frac{\alpha {\bf f}^i} {\sqrt{\gamma}} \frac{\partial\sqrt{\gamma}}
{\partial x^i}.
\end{eqnarray}
\noindent

\subsection{Axisymmetric Coordinate System and Spacetime Variables}

Let us here introduce the notation for the spacetime variables,
writing down the particular expressions used in axisymmetry from
the general case considered in section IIA. Explicit details about
this axisymmetric coordinate choice can be found in~\cite{Brandt94b}.
The metric variables are given by:
\begin{mathletters}
\begin{eqnarray}
\gamma_{ij} & = &\left( \begin{array}{ccc}
\gamma_{\eta\eta} & \gamma_{\eta\theta} & \gamma_{\eta\phi}\\
\gamma_{\eta\theta} & \gamma_{\theta\theta} & \gamma_{\theta\phi}
\\
\gamma_{\eta\phi} & \gamma_{\theta\phi} & \gamma_{\phi\phi}
\end{array} \right) \nonumber\\
&=&  \Psi^4 \left( \begin{array}{ccc}
A & C & E \sin^2\theta \\
C & B & F \sin\theta \\
E \sin^2\theta & F \sin\theta & D \sin^2\theta
\end{array} \right)
\label{threemetric}
\end{eqnarray}
and
\begin{eqnarray}
K_{ij} & = & \Psi^4 H_{ij} \nonumber\\
&=& \Psi^4 \left( \begin{array}{ccc}
H_A & H_C & H_E \sin^2\theta \\
H_C & H_B & H_F \sin\theta \\
H_E \sin^2\theta & H_F \sin\theta & H_D \sin^2\theta
\end{array} \right).
\end{eqnarray}
\end{mathletters}

In these expressions $\eta$ is a logarithmic radial coordinate, and
($\theta$,$\phi$) are the usual spherical angular coordinates. The relation
between $\eta$ and the standard radial coordinates used for
Schwarzschild and Kerr black holes is given by
\begin{equation}
\hat{r}=\frac{\sqrt{M^2-a^2}}{2}e^{\eta}
\end{equation}
\noindent
and
\begin{equation}
r=\hat{r}\left(1+\frac{M+a}{2\hat{r}}\right)
         \left(1+\frac{M-a}{2\hat{r}}\right)
\end{equation}
\noindent
where $M$ is the mass of the black hole, $a$ its angular momentum
per unit mass, $\hat{r}$ a generalization of the Schwarzschild isotropic
radius and $r$ is the usual Boyer-Lindquist radial coordinate.  This
logarithmic coordinate allows one to impose outer boundary conditions at
very large values of the standard radial coordinate.
The conformal factor $\Psi$ is determined on the initial slice, as
explained in section \ref{initdat} below. Since we do not use it
as a dynamical variable, it remains fixed in time afterwards.

\subsection{Boundary Conditions}
Conceptually, the computational domain consists of the region between
two nested spheres, the throat of a black hole and a constant radius
that is very far away (several hundred ADM masses).  At the throat
there is an {\it isometry} condition, which says that all variables
(the metric and gauge variables, the matter fields, etc) at
the interior of the throat can be calculated from their exterior values.

Across the throat of the black hole,
labeled by $\eta=0$, we can demand the condition that the spacetime has
the same geometry for positive $\eta$ as for negative $\eta$. This
condition will build a wormhole into our spacetime. If we formulate
the symmetry correctly, we will obtain simple boundary conditions for
the throat that apply not only initially, but throughout the evolution.
This boundary condition may be stated in one of two ways (in
axisymmetry) to allow for different slicing conditions. Each choice
must result in a symmetric $K_{\eta\phi}$ and an anti-symmetric
$K_{\eta\theta}$ to be consistent with the Kerr solution. Thus, the
form of the isometry condition must be different for spacetimes sliced
with a symmetric or an anti-symmetric lapse.
For an anti-symmetric lapse the condition is 
$\eta \rightarrow -\eta$ while for a
symmetric lapse the condition reads $(\eta,\phi)\rightarrow (-\eta,-\phi)$.

If a lapse that is anti-symmetric across the throat is desired, the 
metric elements with a single $\eta$ index are anti-symmetric across 
the throat, while those with zero or two indices are symmetric. The 
extrinsic curvature components have the opposite symmetry of their 
corresponding metric elements. The shift $\beta^\eta$ is anti-symmetric 
across the throat, while all other shifts are symmetric. 

The scalar matter fields, e.g., density or pressure, 
are simply symmetric across
the throat, producing the same value at $-\eta$ as at $+\eta$.
The momentum fields,
$S^i$ are proportional to the shift divided by the lapse ($\beta^i/\alpha$)
and therefore have the same symmetry as this quantity.

If a symmetric lapse is desired, the metric elements with a single
$\eta$ index or single $\phi$ index (but not both) will be
anti-symmetric at the throat and all others will be symmetric.
The extrinsic curvature components will have the same symmetries as their
corresponding metric elements. The $\beta^\eta$ and $\beta^\phi$
shifts will be anti-symmetric, and the $\beta^\theta$ shift will be
symmetric.  With these symmetries enforced, the initial data and all
subsequent time slices will be isometric across the throat.  One can
verify that all Einstein equations respect these symmetries during the
evolution if they are satisfied initially.

In the $\theta$-direction we can either evolve the whole region from
$\theta=0$ to $\theta=\pi$ or we can use an equatorial plane symmetry
to increase the effective resolution of our simulation.

All metric elements, extrinsic curvature components, shift components,
and the $S^i$ momentum components,
with a single $\theta$ index are anti-symmetric across the 
symmetry axis. The remainder fields
are symmetric.  If we are not using an equatorial plane
symmetry this condition applies both at the axis $\theta=0$ and at
$\theta=\pi$.

At the equator ($\theta=\pi/2$)
there are two possible symmetries, the Kerr symmetry and the ``cosmic
screw'' symmetry. For the Kerr
symmetry, $\theta \rightarrow \pi-\theta$, all metric components,
extrinsic curvature components or shifts with a single $\theta$ index
are anti-symmetric.  The remainder fields
are symmetric. For the cosmic-screw
type boundary conditions the symmetry at the equator is
$(\theta,\phi)\rightarrow(\pi-\theta,-\phi)$ and those metric
elements, extrinsic curvature components, or shift components that
contain one $\theta$ or one $\phi$ index (but not both) are
anti-symmetric.  The remainder are symmetric.

Finally, at the outer boundary a Robin condition is used for 
$\Psi$~\cite{Brandt94a}. This
condition gives the correct asymptotic behavior in the conformal
factor to order $r^{-2}$.  For the metric given in the form
(\ref{threemetric})  $\Psi$ has the form 
\begin{eqnarray}
\Psi = e^{\eta/2}+ (m/2)
e^{-\eta/2}+... 
\end{eqnarray}
\noindent
and therefore obeys the differential equation $\Psi+2
\partial_\eta \Psi = 2 e^{\eta/2}$.  The conformal factor is always
symmetric at the throat, axis, and equator.

\subsection{Gauge Conditions}
\label{gauge}
Following Ref.~\cite{Brandt94b}, we will utilize the gauge freedom 
provided by the shift vector to reduce the number of spacetime 
variables that are evolved.  Two of the shift components are used to 
eliminate the off-diagonal metric variable $C$ and one shift component 
is used to eliminate the off-diagonal metric element$E$.  We note that 
this leaves one degree of gauge freedom unexploited.  Also, due to 
the presence of both even- and odd-parity gravitational wave modes 
(or in other language, the ``plus'' and ``cross'' modes), the metric 
cannot be made completely diagonal.  The $\gamma_{\theta\phi}$ component 
$F$ carries information about the odd-parity wave modes that must be 
accounted for when rotational effects are included.

As for the lapse choice, the code uses the time honored maximal
slicing~\cite{Estabrook73}(that is, one of most commonly used slicing
conditions in evolutions of black holes to date). The singularity
avoiding properties of this slicing are characterized by the
appearance of a limit surface at a distance from the black hole
singularity that is dependent on the angular
momentum~\cite{Eardley79,Duncan85}, and, to a minor extent, on the form of
gravity waves contained within the spacetime~\cite{Brandt94a}. 

Maximal slicing is derived from the freely imposable condition that 
the trace of the extrinsic curvature should vanish throughout the 
evolution.  We note that the Kerr solution in standard form is already 
maximally sliced with antisymmetric lapse, i.e. one which has the 
negative isometry sign across the isometry surface going into the 
black hole.  Setting tr$K=0$ in the evolution for tr$K$ gives
\begin{equation}
-\nabla^a \nabla_a \alpha+\alpha\left( K_{ab} K^{ab}+
4\pi\left(\rho_E+S\right)\right)=0.\label{trkz}
\end{equation}
This elliptic equation is solved numerically on each time slice during the
evolution using a multigrid solver.
This elliptic equation solver~\cite{Schaffer92}
is a semi-coarsening multigrid solver
which does tridiagonal solves along the radial direction.  It has
proved quite robust and reliable in our numerical work to date.

For the purposes of our numerical evolution, the
previous considerations leave the metric 
variables $A$, $B$, $D$, $F$, and all six $H$'s as dynamical variables 
to be evolved.  The various factors of sin$\theta$ are included in the 
definitions to explicitly account for some of the behavior of the 
metric variables near the axis of symmetry and the equator.  

The condition on the shift used in our evolutions is, as in previous work
on vacuum black hole spacetimes, $\gamma_{\eta\theta}=0$
and $\gamma_{\eta\phi}=0$. This choice simplifies the numerical
equations and stabilizes the code.
Since the Kerr shift allows the
stationary rotating black hole metric to be manifestly time
independent, one expects that for the dynamical case a similar shift
will be helpful, and our procedure recovers the Kerr shift given the
Kerr lapse, metric, and extrinsic curvature.

We construct the shift condition by means of the evolution equations
for the two metric variables we are setting to zero.
Let us now consider how to implement the condition $\gamma_{\eta\theta}=0$,
$\gamma_{\eta\phi}=0$ -- or, in terms of our variables, $C=0$ and $E=0$.
The relevant metric evolution equations are:
\begin{mathletters}
\begin{eqnarray}
\partial_t C =&0&=-2 \alpha H_C +
A \partial_\eta \beta^\theta +B \partial_\theta \beta^\eta + F
\sin\theta \partial_\eta \beta^\phi \\
\partial_t E =&0&=-2 \alpha H_E +
D \partial_\eta \beta^\phi
+F \partial_\eta \beta^\theta/\sin\theta.\label{evoHE}
\end{eqnarray}
\end{mathletters} \\
These equations can be combined to produce a single equation involving
$\beta^{\eta}$ and $\beta^{\theta}$:
\begin{equation}
2 \alpha \left(H_C-\frac{F\sin\theta}{D} H_E \right) =
A \partial_\eta \beta^\theta+\left( B - \frac{F^2}{D} \right)
\partial_\theta \beta^\eta.
\end{equation}

We can solve this equation by introducing an auxiliary function
$\Omega$ through the definitions:
\begin{mathletters}
\begin{eqnarray}
\beta^\eta &=& \partial_\theta \Omega,\\
\beta^\theta &=& \partial_\eta \Omega,
\end{eqnarray}\label{defshift}
\end{mathletters} \\
(following Ref.~\cite{Bernstein94a}), producing an elliptic equation 
for the function $\Omega$:
\begin{equation}
2 \alpha \left(H_C-\frac{F\sin\theta}{D} H_E \right) =
A \partial_\eta^2 \Omega+\left( B - \frac{F^2}{D} \right)
\partial_\theta^2 \Omega.
\end{equation}
This equation is then solved by finite differencing using a numerical
elliptic equation solver.  The solution
$\Omega$ is then differentiated by centered derivatives to recover the
shift components $\beta^\eta$ and $\beta^\theta$ according to
Eqs.~(\ref{defshift}).  In practice, these shifts remain fairly small
during the evolution.  Their main function is to suppress the axis
instability, as noted in Ref.~\cite{Bernstein94a} where a similar
shift was used.

Once $\Omega$ is known, $\beta^\phi$ can be calculated by integrating
Eq. (\ref{evoHE}):
\begin{equation}
\beta^\phi=\beta^\phi_{(Kerr)}+
\int_{\eta_{max}}^{\eta} d\eta \left( 2 \alpha H_E
- F\partial_\eta \beta^\theta/\sin\theta \right)/D.
\end{equation}
Only one boundary condition needs to be set (the outer boundary
condition is most convenient), and it is generally set equal to the
Kerr value.  The inner boundary condition, that $\beta^\phi$ must be
symmetric across the throat, is guaranteed by Eq. (\ref{evoHE}).  This
shift component is needed to keep the coordinates from becoming
``tangled up'' as they are dragged around by the rotating hole.
Without such a shift the coordinates would rotate, leading to metric
shear~\cite{Smarr78a}.  This shift component, $\beta^\phi$, is
typically larger than $\beta^\eta$ or $\beta^\theta$.

\subsection{ADM Mass}
Within the axisymmetric coordinate system defined before, 
the ADM mass and angular momentum about
the $z$-axis are defined to be~\cite{Omurchadha74}
\begin{mathletters}
\begin{eqnarray}
M_{ADM} &=& -\frac{1}{2\pi}\oint_S \nabla_a (\Psi e^{-\eta/2}) dS^a
\label{admmass} \\
P_a &=& \frac{1}{8\pi} \oint_S
\left(H_a^b-\gamma_a^b H\right) dS_b.
\label{admp}
\end{eqnarray}
\end{mathletters} \\
In terms of the variables defined in this paper these expressions yield
\begin{mathletters}
\begin{eqnarray}
 M_{ADM} & = & -\int_0^\pi e^{\eta/2} \left(
\partial_\eta \Psi-
\Psi/2 \right) \sin\theta d\theta \label{admmass2},\\
J & = & P_{\phi} = \frac{1}{4} \int_0^\pi \Psi^6 H_E \sqrt{\frac{B\, D}{A}}
\sin^3\theta d\theta
\label{angularmom}
\end{eqnarray}
\end{mathletters} \\
Because of this, the variable $H_E$ is extremely important.  It
determines whether or not angular momentum is present in the spacetime.

Although the ADM mass is defined strictly only at spatial infinity 
$I^0$, in practice we evaluate it at the edge of the spatial grid.  As 
we use a logarithmic radial coordinate $\eta$, this is in the 
asymptotic regime.  While the angular momentum is, in principle, also 
measured at $I^0$, the presence of the azimuthal Killing vector makes 
it possible to evaluate $J$ at any radius.  However, unlike the vacuum 
case, the presence of matter does make it possible to transfer angular 
momentum through the motion of matter if there is angular momentum 
present in the matter initially.  In all simulations in this paper, 
however, the angular momentum of the matter field is initially zero 
(i.e. $S_{\phi} = 0$ by choice; note that this does {\em not} mean 
that $v_{\phi}$ is zero, because that may include rotation of the coordinates).
Therefore we expect that integration 
at any radius will yield $J$.  We monitor this quantity during our 
simulations and use it as a test of the accuracy of our code.

\section{Initial value problem with matter fields}
\label{initdat}

Because the Einstein equations require initial values of the metric
and extrinsic curvature which satisfy constraints (see 
Eqs.~(\ref{hamconstrnt})
and (\ref{momconstrnt})), and because these constraints consist of four
coupled non-linear partial differential equations, obtaining a good
starting point for an evolution requires a special technique -- a conformal
decomposition of the hydrodynamic and spacetime variables.

The basics of this conformal decomposition of
the initial data problem was first given by
Lichnerowicz~\cite{Lichnerowicz44} and later
elaborated by Bowen and York~\cite{Bowen80}.
Essentially, if one makes the following ansatz
\begin{mathletters}
\begin{eqnarray}
\gamma_{ij} &=& \Psi^4 \hat{\gamma}_{ij} \\
\rho_E &=& \Psi^{-8} \hat{\rho}_E \\
S_i &=& \Psi^{-6} \hat{S}_i \\
K_{ij} &=& \Psi^{-2} \hat{H}_{ij}
\end{eqnarray}
\end{mathletters} \\
and if a trace-free extrinsic
curvature is used,
then the Hamiltonian constraint decouples from the momentum
constraints, greatly simplifying the problem.

Next, one normally assumes a form for the conformal metric, typically
that it is flat,
and then constructs a solution to the extrinsic
curvature.  Since this amounts to solving three equations for six
unknowns it is customary to decompose the extrinsic curvature into
a vector potential, $w_j$, (see~\cite{York79}) thereby reducing the number of
unknowns to three:
\begin{eqnarray}
\hat{H}_{ij} &=& 2 \nabla_{(i} w_{j)}-2/3 \nabla_m w^m \gamma_{ij}.
\end{eqnarray}
Standard solutions to this equation with appropriate asymptotic behavior
for linear or angular momentum have been obtained. See~\cite{Bowen80} where
solutions for single boosted or rotating black holes are described.

Another simple solution to the momentum constraints can be found by simply 
inventing a form
for $w^i$, plugging it into the extrinsic curvature solution 
and then solving for the matter current term
\begin{eqnarray}
\hat{S}_i &=& \frac{1}{8 \pi}\nabla^j \hat{H}_{ij}.
\end{eqnarray}

Because the momentum constraint equations are linear in $\hat{H}_{ij}$ it
is possible to combine them with the matter current solution obtained by
inventing a $w^i$ simply by adding them together.  One can add the form
of the extrinsic curvature for a rotating black hole to the simple method
described above, for example, to obtain a solution for initially flowing
matter surrounding a rotating black hole.

At this point we have nearly solved the problem.  We have only two quantities
left to determine: the conformal factor $\Psi$ and the conformal mass
energy density $\hat \rho_E$.  We shall consider $\hat \rho_E$ first.
Our primary consideration in
choosing $\hat \rho_E$ will be to avoid unrealistic
configurations of matter, i.e., matter which does not obey
the energy condition
\begin{equation}
S_a S^a \le \rho_E^2
\end{equation}
and secondarily to find a solution which has matter flowing with a certain
desired speed and along a chosen direction.

For dust, the total velocity of the system can be calculated (thanks to the
choice of the conformal factor made above) by the formula,
\begin{equation}
\frac{ S_i S^i }{\rho_E^2} =
\frac{\hat{S}_i \hat{S}^i}{\hat{\rho}_E^2} = v_i v^i.
\end{equation}
This is guaranteed to be less than one if
we specify the conformal
energy density in the following manner:
\begin{equation}
\hat{\rho}_E = \sqrt{\hat{S}_i \hat{S}^j}+f_E.
\end{equation}
Here $f_E$ is an arbitrarily chosen function that 
respects the boundary conditions
of our problem, and is everywhere greater than zero.  The larger the value of
$f_E$ the slower the matter will move.

The situation is slightly different for the case of a perfect fluid.
In this case we cannot know the precise value of the velocity until we
have completely solved the problem.
At this point the only unsolved quantity that remains is the conformal
factor, $\Psi$, and the only unsolved equation is the Hamiltonian
constraint.  We will solve the
Hamiltonian constraint (a non-linear elliptic equation)
for the conformal factor numerically using the
multigrid solver mentioned above.
Once this is done we can use
Eq.~(\ref{defTSE}) and Eq.~(\ref{TrhouP}) to obtain the following
relations:
\begin{eqnarray}
S^i S_i &=& \left(\rho h W\right)^2 \left(W^2-1\right)\label{SS}\\
\rho_E &=& \rho h W^2 - P \label{Eeq}.
\end{eqnarray}
We can then solve numerically these equations
for $\rho$ and $W$ assuming a polytropic equation of state (i.e., constant
spatial entropy condition)
\begin{eqnarray}
P &=& k \rho^\Gamma \\
\rho h &=& \rho (1+\varepsilon)+P \\
P &=& (\Gamma-1) \rho \varepsilon,
\end{eqnarray}
\noindent
with $\Gamma$ being the (constant) adiabatic index of the fluid.
It is assured by Eq.~(\ref{SS}) that if a solution is found, $v_i v^i < 1$
as in the case of dust considered above.

\section{The Evolution Code}
\label{code}

Now we turn to the description of the numerical code with which
we solve the coupled set of equations presented in 
section~\ref{preliminaries}.
As already mentioned in the introduction, this code is the
result of the merging of two previously existing independent codes.
Each of these two matching pieces was originally developed to solve
only one part of the problem.
The final {\it merged} code
is therefore capable of evolving either a vacuum spacetime, matter flows in
a fixed background or a fully
dynamic spacetime with evolving matter fields.
Due to the ADM
formulation of the Einstein equations, which we are adopting here
for our numerical evolutions, the equations are a mixture of
hyperbolic and elliptic equations. In consequence, it is not 
possible to write the full system as a single, unique system. 
Hence, the code evolves both
fields in an alternate, almost independent, way. 
However, as we shall show below, we can still get second order
convergence to the real solution using this approach.
On the other hand, the alternate integration in time of both systems
of equations
allows one to use different numerical techniques for each one of them,
choosing the best method for each piece. This is in fact what
we do as we will explain later in this section.

As stated previously, 
the hydrodynamical piece of the code makes use of a state-of-the-art modern
high-resolution shock-capturing scheme. These methods are able to handle
discontinuous solutions (e.g., shock waves) 
that could, eventually, develop in
the flow, without using artificial viscosity to damp post-shock oscillations.
To this aim they rely on the so-called 
{\it approximate Riemann solvers}.
The code is capable of using either 
Roe's~\cite{Roe81} or Marquina's~\cite{Donat96,Donat98} methods.
In addition, the code can employ one of several
different cell-reconstruction algorithms to increase the spatial accuracy
of the hydrodynamic evolutions.
Mathematically, these numerical schemes are built upon
the characteristic information of 
the general relativistic hydrodynamics equations. Hence,
the equations have to be written, 
explicitly, as a hyperbolic system of balance laws 
(as Eqs.~(\ref{F}) or~(\ref{hydrosystem})). 

In the hydrodynamic integration of Eq.~(\ref{hydrosystem})
the solution is updated in time, from time $t^n$ to time $t^{n+1}$,
according to the following canonical conservative algorithm
(written in 1D for simplicity)
\begin{eqnarray}
    {\mathbf u}_{j}^{n+1}={\mathbf u}_{j}^{n}-\frac{\Delta t}{\Delta x}
    (\widehat{{\mathbf f}}_{j+1/2}-\widehat{{\mathbf f}}_{j-1/2}) +
    \Delta t {\mathbf s}_j,
  \end{eqnarray}
\noindent
where index $j$ indicates cell centers and indices $j\pm 1/2$ indicate
cell interfaces.
In particular, and in order to increase the temporal order of the 
scheme,
a high-order (typically second or third) 
monotonicity preserving Runge-Kutta method~\cite{Shu88} 
is used to update the 
solution in time. For a third-order scheme the algorithm looks
like this:
\begin{mathletters}
\begin{eqnarray}
{\mathbf u}_{j}^{1} &=& {\mathbf u}_{j}^{n} + \Delta t L({\mathbf u}_{j}^{n})
\\
{\mathbf u}_{j}^{2} &=& {\mathbf u}_{j}^{n} + \frac{1}{4} 
 \Delta t L({\mathbf u}_{j}^{n}) + \frac{1}{4}
 \Delta t L({\mathbf u}_{j}^{1})
\\
{\mathbf u}_{j}^{n+1} &=& {\mathbf u}_{j}^{n} + \frac{1}{6}
 \Delta t L({\mathbf u}_{j}^{n}) + \frac{1}{6}
 \Delta t L({\mathbf u}_{j}^{1}) + \frac{2}{3}
 \Delta t L({\mathbf u}_{j}^{2}),
\end{eqnarray}
\noindent
\end{mathletters} \\
with
\begin{eqnarray}
L = -\frac{(\widehat{{\mathbf f}}_{j+1/2}-\widehat{{\mathbf f}}_{j-1/2})}
    {\Delta x} + {\mathbf s}_j.
\end{eqnarray}

A monotonic linear reconstruction of
the cell centered values of the primitive variables provides second-order
accuracy in space~\cite{vanLeer79}. 
Finally, the {\it numerical fluxes} across interfaces, 
$\hat{\mathbf f}_{j\pm 1/2}$, are
calculated, in the frame of the local characteristic approach, according to
\begin{eqnarray}
\hat{\bf f}_{j\pm 1/2} &=& \frac{1}{2}
({\bf f}_{j\pm 1/2}({\bf w}_{\rm R}) +
 {\bf f}_{j\pm 1/2}({\bf w}_{\rm L}) \nonumber\\
 &-& \sum_{n=1}^5 |\widetilde{\lambda}_{(n)} |
 \Delta \widetilde{\omega}_{(n)}
\widetilde{\bf r}_{(n),{j\pm 1/2}} )
\label{nflux}
\end{eqnarray}

\noindent
where ${\bf w}_{\rm L}$ and ${\bf w}_{\rm R}$ represent the values of the
{\it primitive variables}
at the left and right sides, respectively, of the
corresponding interface. This state vector is defined as
${\bf w}=(\rho,v_i,\varepsilon)$.
In addition, $\{\widetilde{\lambda}_n, \widetilde{\bf r}_n\}_{n=1,..5}$ are,
respectively, the eigenvalues and right-eigenvectors of the Jacobian
matrix of the system calculated at the interfaces from
${\bf w}_{\rm L}$ and ${\bf w}_{\rm R}$.
Explicit general expressions can be found in~\cite{Banyuls97}.
Finally, the quantities $\{\Delta \widetilde{\omega}_n\}_{n=1,..5}$, the
jumps of the characteristic variables across each characteristic field,
are obtained from
\begin{equation}
{\bf u}({\bf w}_{\rm R})-{\bf u}({\bf w}_{\rm L}) =
\sum_{n=1}^5 \Delta \widetilde{\omega}_n \widetilde{\bf r}_n.
\end{equation}
\noindent
The tilde in some of the previous quantities indicates
averaged quantities at the cell interfaces, algebraically computed from
${\bf w}_{\rm L}$ and ${\bf w}_{\rm R}$.
Further information can be obtained in~\cite{Banyuls97} and~\cite{Font94}. 

For the spacetime part of the code we do not use a shock-capturing scheme.
In our studies to date the metric variables are generally smooth 
so this presents no problem.
However, in principle, any hydrodynamical shock can affect
the spacetime part in terms containing first (time) derivatives of
the extrinsic curvature components, as shown by Eq.~(\ref{excurv}).
Hence, a shock-capturing scheme could still be a good choice.
In addition, due to the slicing conditions and
as an attempt to avoid any singularity appearing on the spacetime, the
metric quantities can (and in fact do) develop large gradients (see, e.g.,
\cite{Brandt94b}). In practice, these closely
resemble the steep ones which characterize any real 
hydrodynamical shock. The use of 
shock-capturing schemes for the spacetime is, however, restricted to the
recently developed hyperbolic formulations of the Einstein equations
(see~\cite{Bona96a,Reula98a,Bona98b} and references
therein). We are also presently 
exploring the use of such advanced methods on the fully coupled 
equations in hyperbolic form~\cite{Font97}, \cite{Aloy98a}, \cite{Font98b}.

The spacetime variables are
evolved for the full set of Einstein equations using an 
explicit second order Runge-Kutta scheme.
Schematically the evolution looks like this:
\begin{mathletters}
\begin{eqnarray}
\tilde{\gamma}^{n+1/2} &=& \gamma^n+\frac{1}{2}\Delta t \dot{\gamma} \left(
K^{n}, \gamma^{n} \right) \\
\tilde{K}^{n+1/2} &=& K^n+\frac{1}{2}\Delta t \dot{K} \left(
K^{n}, \gamma^{n} \right) \\
\gamma^{n+1} &=& \gamma^n+\Delta t \dot{\gamma} \left(
\tilde{K}^{n+1/2}, \tilde{\gamma}^{n+1/2} \right) \\
K^{n+1} &=& K^n+\Delta t \dot{K} \left(
\tilde{K}^{n+1/2}, \tilde{\gamma}^{n+1/2} \right) 
\end{eqnarray}
\end{mathletters} 
Spatial derivatives needed in the above equations are calculated
using centered, second order finite differencing.   This is a different
scheme than we have employed in the past~\cite{Brandt94b}.  We switched
the procedure in order to simplify the alignment of time levels.  This
different evolution procedure has proved to be as stable and reliable
in our simulations as our previous scheme.

\section{Analysis Tools}
\label{sec:analysis}
Now we discuss the use of several tools we have developed to extract the 
physics from a fully relativistic simulation of dynamic spacetimes 
involving black holes.  Unlike the standard cases involving 
hydrodynamics, where density, pressure and fluid velocities have well 
defined meanings, in a dynamic spacetime, the metric functions 
themselves, which are the functions actually evolved, do not carry 
direct physical or geometric meaning.  Rather, they are 
related to the coordinate system and gauge in which the simulations 
are carried out.  Therefore, physically relevant information about 
important quantities such as gravitational radiation or the mass of the 
black hole must be derived from the metric functions.

\subsection{Waveform extraction}
The gravitational radiation emitted is one of the most important 
quantities of interest in many astrophysical processes.  However, the 
radiation is generated in regions of strong and dynamic gravitational 
fields, and then propagated to regions far away where it can
be detected.  We take the approach of computing the evolution of the 
fields in a fully nonlinear way, while analyzing the radiation 
emitted in the regions where the system can be considered as a 
perturbation propagating on a fixed background.

Under these conditions, one can appeal to the well developed theory of 
black hole perturbations.  In this case one identifies certain 
perturbed metric quantities that evolve according to wave equations on 
the black hole background.  However, the perturbed metric functions 
are also dependent of the gauge in which they are computed.  
Fortunately, there is a {\em gauge-invariant} prescription for 
isolating wave modes on black hole backgrounds, developed first by 
Moncrief~\cite{Moncrief74}.  The basic idea is that although the first 
order perturbed metric functions transform under first order 
coordinate transformations (gauge transformations), one can identify 
certain linear combinations of these functions that are invariant 
under such transformations.  These gauge-invariant functions are 
clearly more directly related to true physics (which does not depend 
on coordinate systems), and in fact these functions obey the wave 
equations describing gravitational wave modes in linear theory.  There 
are two independent wave modes, even- and odd-parity, corresponding to 
the two degrees of freedom, or polarization modes, of the waves.  The 
metric used in this work allows both modes, as discussed in 
Sec.~\ref{gauge} above.

In principle, one can consider a full set of gauge-invariants 
including both matter and gravitational wave fields, but our interest 
here is in evolving the hydrodynamics and spacetime within the fully 
nonlinear theory, and simply extracting the gravitational wave 
information assuming a vacuum region away from the black hole.  A 
waveform extraction procedure has been developed that allows one to 
process the metric and to identify the wave modes.  The gravitational 
wave function (often called the ``Zerilli function'' for even-parity 
or the ``Regge-Wheeler function'' for odd-parity) can be computed by 
writing the metric as the sum of a background black hole part and a 
perturbation: $g_{\alpha\beta}=\stackrel{o}{g}_{\alpha\beta}+h_{\alpha 
\beta}$, where the perturbation $h_{\alpha\beta}$ is expanded in 
spherical harmonics and their tensor generalizations.  To compute the 
elements of $h_{\alpha\beta}$ in a numerical simulation, one 
integrates the numerically evolved metric components $g_{\alpha\beta}$ 
against appropriate spherical harmonics over a coordinate 2--sphere 
surrounding the black hole.  The resulting functions can then be 
combined in a gauge-invariant way, following the prescription given by 
Moncrief\cite{Moncrief74}, leading to the gravitational wave 
functions denoted by $\psi$.
Then for each $\ell -m$ mode, one can extract the 
waveforms of gravitational waves as they pass a ``detector'' at some 
fixed radius.  This procedure has been described in detail for the 
case in this paper in~\cite{Brandt94a,Brandt94b,Brandt94c}, and more 
generally in Refs.~\cite{Allen97a,Camarda97c}.

\subsection{Horizon tools}

We now turn to the topic of black hole horizons in numerical 
relativity and their application to the spacetimes considered in this 
paper.  Horizons can be used in various ways to analyze the physics 
of a black hole system, and also as a check on the accuracy of the 
spacetime.  We briefly discuss these issues below, and then apply 
horizon finders in testbed simulations below.  

There are two definitions of black hole horizons of interest to us: 
(a) The {\em event} horizon (EH) is the most commonly encountered 
term, defined loosely as the (closed, 2D) surface that separates those 
light rays that can escape the black hole's gravitational pull from 
those light rays that cannot.  Exactly on this critical surface are 
light rays that never fall in to the black hole, and never escape to 
infinity.  This surface is impossible to find on a given time slice, 
as photons that appear to be propagating (expanding) away from the 
black hole at one time may later find themselves falling back into the 
black hole if more mass-energy falls in, increasing its gravitational 
pull.  Thus, the EH is generally an expanding surface composed of 
photons that will eventually find themselves trapped.  Hence, locating 
the {\em event} horizon requires an entire evolution of a black hole 
spacetime.  Methods do exist for finding event 
horizons~\cite{Anninos94f,Libson94a,Hughes94a,Masso95a} and can be 
applied to spacetimes like those considered here, but we shall employ
that application in future work.  (b) The {\em apparent} horizon 
(AH) is defined loosely as the (outermost, closed, 2D) surface on 
which all outgoing photons normal to the surface are not {\em 
instantaneously} expanding away or converging towards each other: they 
have zero expansion.  In stationary black hole systems, where no 
mass-energy will fall into the black hole, the AH and EH coincide, but 
generally the AH lies inside the EH.  It is a convenient surface to 
locate on a given time slice, since one only needs to find a closed 
surface such that the expansion of all outgoing photon bundles have 
zero expansion.

The expansion $\Theta$ of a congruence of null rays moving in the 
outward normal direction to a closed surface can be shown to 
be~\cite{York89}
\begin{equation}
\Theta = \nabla_i s^i + K_{ij} s^i s^j - {\rm tr} K ,
\label{eqn:expansion}
\end{equation}
where $\nabla_i$ is the covariant derivative associated with the
3-metric $\gamma_{ij}$, $s^{i}$ is the normal vector to the surface, 
$K_{ij}$ is the extrinsic curvature of the time slice, and ${\rm tr} 
K$ is its trace. An AH is then the outermost surface such that
\begin{equation}
\Theta = 0.
\label{eqn:horizon1}
\end{equation}
This equation is not affected by the presence of matter, since it is 
purely geometric in nature, and so we can use standard horizon 
finders without modification for our current non-vacuum spacetimes.

The key is to find a surface with normal vector $s^{i}$ satisfying 
this equation.  There are many methods designed by now to locate 
apparent horizons in 2D, and we shall use the horizon finder developed 
for the black hole code used in this work, described in 
Ref.~\cite{Brandt94a,Brandt94b,Brandt94c}.  We refer the reader to 
those papers for details of the algorithm.  In those references it was 
shown how one can use the dynamics and geometry of the AH surface to 
study the physics of dynamic, rotating vacuum black hole spacetimes.  

For the purposes of this paper, the most important physical quantity 
to be used is the area $A$ of the apparent horizon.  From this we compute 
an effective mass $M_{AH}$ of the black hole, defined in
axisymmetry~\cite{Christodoulou70} by
\begin{equation}
	M_{AH}^2 = \frac{A}{16\pi}+\frac{4 \pi J^2}{A}\label{MAH}.
\end{equation}
We know that as the black hole approaches a stationary state at late 
times, the apparent and event horizons will coincide, and in that case 
the mass of the black hole is rigorously defined by the above formula.  
One can use this black hole mass to study energy accounting: the total 
ADM mass of the spacetime, defined by Eq.~(\ref{admmass}) above, should be 
equal to the final black hole mass plus any mass energy carried away 
by gravitational waves to infinity.  This gives a powerful check on 
the overall global accuracy and consistency of the code, testing 
several crucial and independent aspects of the code and physics 
extraction models.  These were very powerful tests in the vacuum 
case~\cite{Brandt94a,Brandt94b,Brandt94c}.  We shall apply such tests 
below and show to what extent they are useful in these spacetimes with 
matter flows onto black holes.

\section{Tests and Calculations}
\label{results}

In this section we present a summary of consistency checks and
testbed computations that the code has successfully passed.
We have first verified that the code reproduces previous results 
when either the matter fields are set to zero,
or the evolution of the spacetime is turned off.  We 
start by describing convergence tests of the code, for some particular 
initial matter configurations, to see the order of the method.  This 
is a necessary requirement as no exact solutions to compare with 
exist. We then move on to consider simulations of the
spherical accretion of dust and perfect fluid matter onto a dynamic
black hole. In this application we show the differences between integrating
the hydrodynamical equations fully coupled to the spacetime or in the
simplified case of
a sliced spacetime which does not react to the presence of matter. As
expected, for sufficiently low energy density initial distributions, the 
spacetime evolution is totally unaffected by the matter content.
Finally, we explore a few axisymmetric spacetimes and make some
preliminary explorations of matter accreting onto
both rotating and nonrotating black holes.

\subsection{Convergence tests}

We have performed a series of convergence tests on our code.
We measure convergence along the line $\theta = \pi/4$ for
the constraints and other functions. Because we do not
have data placed along this value of $\theta$ we interpolate it from
our existing data using a third order interpolation scheme.  The
convergence rate of the Hamiltonian and momentum constraints is
based on data at two resolutions and the assumption that the true
value is zero, and for any other quantity the convergence
is calculated by comparing
results obtained at three resolutions (keeping the ratio
$\Delta \eta/\Delta \theta$ fixed) in a similar manner as reported
in~\cite{Bernstein93b}. In all cases, we assume that our
functions are converging according to the formula
\begin{equation}
x_n = x_{True} + k (n \Delta \eta)^\sigma
\end{equation}
where $k$ is a constant and $n$ is either $1$, $2$, or $4$ depending
on whether we are using high, medium, or low resolution.
The constraints are thus measured by substituting the numerically
computed constraint for $x_n$, zero for $x_{True}$ and the (radial) grid
resolution for $\Delta \eta$.  The convergence of the constraints
is given by
\begin{equation}
\sigma = \log\left|\frac{x_2}{x_1}\right|/\log 2.
\end{equation}
\noindent
When we are not converging toward a known answer, for example when
testing the convergence of the values of the field variables for the
matter and the spacetime metric, we use the standard three point
convergence formula
\begin{equation}
\sigma = \log\left|\frac{x_4-x_2}{x_2-x_1}\right|/\log 2.
\end{equation}
In all cases that we show in this paper, we find that the convergence 
rate $\sigma$ is approximately $2$ for most quantities, i.e., the code 
is globally second order accurate.  It is worth mentioning that the 
two starting codes from which the present code has been built upon, 
that is the matter and the vacuum spacetime codes, were also, 
originally, second order accurate.  The coupling of them both into a 
unique code has not diminished the final convergence order.


For a sample case, analyzed in detail in Section~\ref{rotating}
below, we show convergence numbers in Table~\ref{tab:converge} and
two plots of the constraint violation
(See Figs.~\ref{fig:ham}, \ref{fig:mom1}).
The convergence is measured after an evolution time of $6M$.  The
resolutions used for this test are $100\times10$, 
$200\times20$, and $400\times40$ zones, in the radial and
polar directions, respectively. The
parameters describing the conformal density of the initial model are
$(\rho_b,\kappa,\rho_{max},\eta_0,n)=(0.01,1,3,2,2)$ and the angular
momentum of the spacetime
is $10$ (to understand the meaning of these parameters 
we refer to Eq.~(\ref{shell})).  In the plots, we show the violation
of the Hamiltonian constraint and the $\eta$-component of the momentum
constraint for the medium ($\times 1/4$)
and high resolution runs.  If the equations
are converging at second order, then the two lines in each figure
should be right on top of one another -- as they are.

\subsection{Spherically Symmetric Simulations}
\subsubsection{Dynamically sliced accretion}

We have first tested the hydrodynamical piece of the code against the 
analytic, spherically symmetric, Bondi accretion 
solutions~(\cite{Bondi52}, see also~\cite{Michel72} for its
relativistic extension). We have verified that the code 
reproduces them, both for dust and a perfect fluid, when the 
background metric is kept fixed, to within a few percent for 
evolutions of $100 M_{ADM}$. 

However, in dynamic spacetimes, one will not normally use a slicing 
that maintains a static metric, even if the underlying geometry has 
evolved to a stationary state.  Therefore it is important to develop 
testbeds of known results, like the spherical Bondi accretion solution, 
but in non-analytic 
slicings which are commonly used in numerical relativity.  We shall
employ maximal slicing, but with different boundary conditions than those
found in the stationary Schwarzschild metric.  Therefore 
we compare the evolution of matter in a {\em dynamically} sliced spacetime 
which does not react to the presence of that matter distribution.  In 
other words, the metric variables evolve completely independent of the 
hydrodynamical quantities (just as in vacuum spacetimes) whereas the 
latter evolve in a dynamic spacetime.  This is a new type of testbed 
that we are developing for matter flows accreting onto dynamic,
and dynamically sliced, black holes. 

We now wish to compare the analytic solution to the numerical one obtained in 
this dynamic slicing.  It is nontrivial to do this in general, since one must 
compare invariant quantities at the same spacetime points in the two
systems.  Coordinate values will not suffice, since in the dynamically 
sliced case they are moving through spacetime.
In the original slicing everything
is a function of $r$ only, while in the new slicing everything depends on space
and time, i.e. on the new coordinates
$\bar{r}$ and $\bar{t}$.  The rest-mass density,
for example, is given by $\rho(r)$ in the analytic solution, but in the
new slicing it is an evolving quantity $\rho(\bar{r},\bar{t})$.
Fortunately, it is possible to reconstruct $r$ from $\bar{r}$ and $\bar{t}$
and thus to compare the time-evolved quantity, $\rho(\bar{r},\bar{t})$, with 
the analytic solution, $\rho(r)$. 
The value of $r$ is simply the areal radius
$r = \sqrt{\gamma_{\theta \theta}}$.  
In Fig.~\ref{BondiComp} we see the result of an evolution with a
non-Schwarzschild slicing.
We plot the numerically evolved value of $\rho$ (i.e. $\rho(\bar{r},\bar{t})$)
with a dotted line.
For reference, we also plot the initial value with a solid line.
Note that when $\bar{t}=0$ it
is true that $r=\bar{r}$. 
The line marked with circles ($\rho_{exact}$)
represents the value of $\rho$ as a function of the reconstructed $r$
(i.e. $\rho(r) = \rho(\sqrt{\gamma_{\theta\theta}})$).
The solid line and the line for $\rho_{exact}$ are really 
the same function, therefore, plotted with the areal radius as calculated
at the times $t=0$ and $t=20 M$ respectively.
Clearly the numerically evolved solution
(dotted line) agrees with the analytic ($\rho_{exact}$)
solution to a high accuracy (and the constraints converged to second order).

\subsubsection{Dust accretion}

We now increase the complexity of the problem by ``switching on" the matter
fields appearing in the spacetime equations.
We start showing the effects of a full coupling with the same matter fields
of the previous section. 
Hence, initially, we assume a spherical distribution of dust with zero 
velocity and constant conformal energy density. 
Then, we solve the Hamiltonian and momentum constraint 
equations in order to obtain initial thermodynamic profiles satisfying 
Einstein equations. 


First, we consider a small uniform density, 
$\hat{\rho} = 10^{-2}$, and solve the Hamiltonian constraint at $t=0$, 
which, in turn, modifies the spacetime geometry.
In Fig.~\ref{dynslice1} we show the evolution of the radial metric 
function $A$ up to a final time of $50 M_{ADM}$ in intervals of $5M$.
We use a coarse 
radial grid of $150$ zones and only one angular zone, with the outer 
boundary placed at $\eta=6.5$.  The solid line in this figure
corresponds to the dynamically sliced Schwarzschild evolution discussed
previously (a {\it partly coupled} evolution). Now we find, of course, a very 
different behavior from the standard Bondi accretion, where this 
metric function would remain time independent, which is merely a 
manifestation of the singularity avoiding maximal slicing used.
The evolution of the radial metric 
function $A$ for the fully coupled case
is shown as a dotted line in Fig.~\ref{dynslice1}. Note that for 
such a small density distribution, the effects of the matter 
on the spacetime geometry evolution are negligible.

Now we show the effect of increasing the amount of dust.  
Fig.~\ref{dynslice2} corresponds to $\hat{\rho}=1$. 
Notice first that the peak of 
the metric function $A$ at $t=50 M_{ADM}$ (last curve) is located at a 
coordinate distance $\eta$ slightly larger than 2 regardless the value 
of $\rho$ for the dynamically sliced Schwarzschild evolution (as one 
would expect). However, in the fully coupled 
evolutions, matter falls onto the black hole increasing its radius and 
causing the lapse to collapse more quickly in a larger interior 
region.  As a result of this, the grid stretching effect which creates 
the peak in $A$ occurs, in the high density run,  
at a more distant part of the grid than for the 
low density case. We have also carried out 
simulations with $\hat{\rho}=10^{-3}$.
In this very low density case, as expected,
the dotted and solid lines, or in other words, the fully and partly
coupled evolutions, totally coincide.



Let us now analyze the behavior of some characteristic relevant fields
in these simulations. In the remaining of this section we consider the
$\hat{\rho}=10^{-2}$ case.
Fig.~\ref{sphacc1} depicts the time evolution sequence of the lapse, 
$\alpha$, the conformal radial metric variable $A$, and two 
hydrodynamical variables, the rest-mass density, $\rho$
and the total velocity $v$.  For this last 
quantity we plot its magnitude, multiplied by the sign of 
the radial component of the velocity (i.e. sign($v_{\eta}$) 
$\sqrt{v^{a}v_{a}}$), in order to indicate if the matter is accreting onto 
the hole (negative values) or expanding away (positive values).  
The collapse of 
the lapse at the innermost zones freezes the evolution there at an 
early time, permitting us to avoid the singularity and continue the 
evolution until typical times of $\approx 100M_{ADM}$.  
The radial metric function 
$A$ shows its characteristic peak (a result of tidal forces on the 
grid points) which, at late times, leads to numerical problems in the 
integration of the metric part of the code.  The density of the dust 
grows but rapidly settles in the interior region, due to the collapse 
of the lapse in that part of the grid and, although it is not visible 
in the plot, it continuously but slowly evolves in the outer zones.  
The evolution of the total velocity clearly shows that, initially, the 
matter accelerates rapidly, reaching $\approx 0.8 c$ within the first 
$20M$ of evolution.  However, due to the freezing of the lapse, this 
evolution is slowed and we never see it go past $0.9 c$ even after 
$100M$ of evolution.  If the evolution were continued long enough we 
would expect it to asymptotically approach $c$, which is the free-fall 
velocity at the horizon measured by an Eulerian observer in this 
particular coordinate system.

In Fig.~\ref{sphacc2} we plot the time evolution of the apparent 
horizon mass for three different radial resolutions (150, 300 and 600 
zones).  Correspondingly, in Fig.~\ref{sphacc2_2} we plot the location 
of the apparent horizon.  We only show its location for the most 
resolved run (600 radial zones) as the results are almost identical 
for the lower resolutions.  Note how sensitive the calculation of the 
horizon mass is to knowing its precise location.  At lower 
resolutions, 150 and 300 zones, one sees a rapid, but spurious, growth 
of the horizon due to violation of the Hamiltonian constraint which 
actually causes it to go above the ADM mass within $t=50M$.  In 
Fig.~\ref{sphacc2} we can see that initially $70\%$ of the mass energy 
is contained within the apparent horizon, but by $t=50M$ almost all of 
this has fallen onto the black hole.


\subsubsection{Perfect fluid accretion}

Now let us increase again the complexity of the equations
including the corresponding pressure terms which were absent in the 
dust case.  Comparing to the accretion of pressureless matter, 
additional assumptions must be made in order to assign definite values 
to the density and pressure, as was described in 
section~\ref{initdat}. Furthermore, we make the choice that $P 
\rho^{-\gamma} = constant$ initially.  In Fig.~\ref{sphacc3} we plot the time 
evolution of the same quantities of Fig.~\ref{sphacc1} but now for the 
spherical accretion of a perfect fluid of adiabatic exponent $4/3$.  
As before, we consider a constant conformal mass 
density $\hat{\rho} = 10^{-2}$.
Now, the role of the pressure makes the hydrodynamical 
evolution somewhat different from the pressureless case.  This is 
clearly noticeable in the behavior of the velocity field.  Initially 
all the material is at rest.  The evolution of the velocity proceeds 
very rapidly and, in just $10M$, we can identify part of the material 
falling towards the hole (negative values, using the same convention 
as for the dust case) and part of it {\it expanding} to infinity 
(positive values).  The reason for this is the existence of a high 
pressure distribution of matter surrounding the hole and, hence, a 
pressure force that can, eventually, halt the gravitational collapse.  
The consequence is that part of the material bounces back towards 
larger values of $\eta$.  This is in clear contrast with the dust case 
where no other force exists that can support gravity and the fluid can 
only freely-fall towards the hole.  Finally, in Fig.~\ref{sphacc4} we 
plot the pressure and internal energy density evolution for this run.



\subsection{Axisymmetric Simulations}

We present axisymmetric simulations showing the evolution of
imploding (accreting) extended matter shells onto the black hole.
We also discuss some results concerning the
evolution of matter in rotating spacetimes.  The simulations we
present here aim to show the performance and feasibility of our
numerical tool.  In future work we plan to carry over a detailed
comparative and parametric study of the different astrophysical
scenarios just outlined here.

\subsubsection{Imploding shells of dust}
\label{shellsect}

As the first two-dimensional problem we study the evolution of an 
imploding shell of dust that radially falls towards the black hole.  
This problem has been studied semi-analytically in the 
past~\cite{Nakamura81b}, \cite{Shapiro82} where its astrophysical 
implications in gravitational wave astronomy were 
discussed.  In~\cite{Nakamura81b}, the academic problem of computing 
the gravitational radiation from {\it infinitesimally thin} spheroidal 
sheets falling into black holes was discussed.  In~\cite{Shapiro82}, 
Shapiro and Wasserman extended these computations to {\it finite size} 
shells and pointed out the possible relevance of this process in 
non-spherical gravitational collapse to a black hole and in 
axisymmetric accretion onto a black hole.  The most interesting 
conclusion of these semi-analytic studies was to demonstrate that the 
total energy emitted in gravitational radiation by a non-spherical 
dust cloud falling into a black hole is always less than that for a 
point particle of the same mass falling into the hole.  This 
suppression was explained as a result from interference between waves 
emitted from different parts of the extended object.

This problem has recently been studied numerically 
in~\cite{Papadopoulos98c} with a {\it hybrid} coupled code which 
employs linearized (perturbative) gravity (the Teukolsky equation with 
matter sources~\cite{Teukolsky73}) and fully nonlinear hydrodynamics.  
The hydrodynamical piece of that code coincides with the one we use in 
the present fully non-linear code.  In~\cite{Papadopoulos98c} 
Papadopoulos and Font 
were able to demonstrate, in the linear regime, the suppression 
of gravitational radiation
emission for a class of extended objects.  Moreover, 
they showed that the radiated energy approaches an asymptotic value
as the initial density distribution in the shell is made increasingly 
more compact.

We present here the first numerical results of the fully non-linear 
evolution of an imploding shell of dust.  Our aim is to capture the 
essential features of the problem.  In order to do so we focus on a 
single initial model delaying for a forthcoming work a comparative 
study of the relevant parameter space of the problem (shell mass and 
compactness, shell-black hole separation, etc.).  We are now mainly 
concerned in showing how the presence and evolution of the shell 
triggers the emission of gravitational radiation.

We consider a Gaussian shell of dust whose conformal density distribution
is parameterized by its location, $\eta_0$,
amplitude, $\rho_{max}$ and width, $\kappa$, according to the following
formula
\begin{equation}
\hat\rho=\rho_{b} + \frac{1}{2}\rho_{max}\left(
e^{-\kappa(\eta-\eta_0)^2}+
e^{-\kappa(\eta+\eta_0)^2}
\right)\sin^n\theta,
\label{shell}
\end{equation}
\noindent where $\rho_b$ is the background density.  In particular we 
chose $\rho_b=10^{-2}$, $\rho_{max}=10$, $\kappa=0.5$, $n=4$, and 
$\eta_0=2$.  We employ a grid of $300 \times 25$ zones in $\eta$ and 
$\theta$, respectively.  Then solve the
Hamiltonian constraint, Eq.~\ref{hamconstrnt}.
The angular coordinate runs from $0$ to 
$\pi/2$ and the radial one from $0$ to $7.5$.  The results of the 
simulation are plotted in Fig.~\ref{shell1} for a final time of 
$100M$.  Here we plot $\alpha$, $A$, $\rho$ and $v$.  One can again 
see the characteristic collapse of the lapse in the inner regions and 
the growth of the metric component $A$.  The behavior of the total 
velocity is also similar to that found in the spherical accretion 
problem.  The non-spherical aspect of this simulation is the initial 
distribution of the matter density.  As this initial distribution has 
equatorial plane symmetry we only plot the results in the first 
quadrant.

We plot in Fig.~\ref{shell2} profiles of the density along the axis 
($\theta=0$) and for different times of the evolution.  
Clearly noticeable is the infall of the shell towards the hole 
accompanied by its collapse.  As the shell does not have spherical 
symmetry, this implosion induces the emission of gravitational waves.  
We compute the even-parity $\ell=2$ and $\ell=4$ modes of the emitted 
radiation and they are plotted in Figs.~\ref{shell3} and~\ref{shell4}, 
respectively (odd-parity modes are absent in these non-rotating 
simulations).  We also show in these plots a fit against the first two 
harmonics of the corresponding quasinormal modes.  Clearly, they agree 
quite closely with the wave one expects to see for a black hole with a 
mass comparable to the entire dust-shell-plus-black-hole system.  We 
emphasize that these quasinormal mode fits are not perturbative 
evolutions, as discussed above, but rather comparisons with known 
complex oscillation frequencies of black holes.  The phase and 
amplitude of the two lowest modes are adjusted for a best fit to the 
obtained waveforms.  We also compute the total energy radiated away by 
these two modes.  For the $\ell=2$ we find $2.36 \times 10^{-6}$ and 
for the $\ell=4$ we get $6.35 \times 10^{-9}$.  These values are 
normalized to the ADM mass of the spacetime.


Next, we turn to another technique for studying a 
black hole accreting matter: examining the horizon dynamics.  As 
discussed above, the black hole apparent horizon can give important 
information about the system.  Its area is related to the black hole 
mass, and combining the study of radiation emitted by gravitational 
waves can provide powerful checks on the overall energy accounting of 
the system.  Energy conservation is very important in traditional 
hydrodynamics, but for dynamic spacetimes we must develop new 
techniques to account for radiation and the difficulty of localizing 
energy in general relativity.  In this case one can use several 
completely independent measures of the energy in the system that 
should be related: the final mass $M_{AH}$ of the black hole, after all 
mass-energy has fallen in, will be given by the horizon area through 
Eq.~(\ref{MAH}) and the total energy radiated, $E_{rad}$ can be 
computed from the Zerilli function $\psi$.  In principle these should 
add up to the total ADM mass of the spacetime, $M_{ADM}$, computed at 
large radius.  This has been used extensively in vacuum. We now apply 
these techniques to black holes surrounded by matter.


We consider a black hole plus matter spacetime with a non-flat
background metric, hence inducing a more significant contribution of
radiation energy.  The specific non-flat background metric we employ
is called the Brill wave metric, and it allows us to place a packet of
gravitational wave energy of any desired width at any distance
from the central black hole.
For details of the construction of this metric see~\cite{Brandt94a}.
In Fig.~\ref{account} we show various energy 
measures for this system.  The thick lines shows the mass of the apparent 
horizon as a function of time.  It grows as matter and gravitational 
wave energy fall in rapidly, and then settles down at roughly
$t=20M$.  It is actually still growing slowly here 
because some matter is still falling in, but the amount is negligibly small.
In order to compute the apparent horizon mass to the accuracy
desired for this plot it was necessary
to run at very high resolutions -- in the final case shown the evolution
was performed using 900 radial zones.  As noted previously, this resolution
is not needed to accurately compute the position of the apparent horizon,
nor the energy in the radiation zone.


We also show the ADM mass as a solid thin line, indicating the 
total mass of the spacetime.  In principle, the black hole mass cannot 
exceed this limit, although it does at late times due to numerical 
error.  Finally, we indicate the total radiated energy, computed 
through the Zerilli function, by a dotted thin line.  The distance between 
the solid and dashed lines equals $E_{rad}$.  If all mass energy has 
gone into the black hole, one should see $M_{AH} + E_{rad} = M_{ADM}$.  
Fig.  \ref{account} shows that this is quite closely achieved 
numerically, even though the total energy radiated is 
only $0.0874 M_{ADM}$.  The 
small gap between the energy radiated and the horizon mass is 
attributed to two effects: first, not all matter has actually fallen 
into the horizon by this time, and second, the apparent horizon mass 
will always be less than the event horizon mass in such cases.  
We regard this energy accounting issue 
as an important test and diagnostic of the physics of 
black hole accretion in dynamic black hole spacetimes.  
However, as one can see from Fig.~\ref{account}, our studies indicate
that this test is a very sensitive measure of global error.
The crucial difficulty lies in resolving the peak in the metric
function $A$ that develops near the horizon. Small errors there
translate into large deviations in the area calculation. Apparent horizon
boundary conditions will aid this kind of study greatly.

\subsubsection{Rotating spacetimes}
\label{rotating}

Finally, we present some results concerning the evolution of matter in 
rotating spacetimes.  The way in which these spacetimes are 
constructed and evolved is the same as in~\cite{Brandt94b}.  In a 
rotating spacetime, constant $\eta$ observers along the equator are 
spinning around the black hole.  In consequence, their fall through 
the horizon is slower and grid stretching is less near the equator.  
This is noticeable in the plots we show below.  For the rotating cases 
below we generally choose a lapse which is symmetric across the throat, 
although in principle either choice (symmetric or
antisymmetric) is possible.

To show the behavior of the code when rotation is present
we choose the same shell as in the previous section. The initial data
set for the spacetime is based on a rotating Bowen and York~\cite{Bowen80}
black hole. We use this construction instead of Kerr for 
the sake of simplicity. We should note, however, that it contains 
radiation in 
the initial slice due to the construction of the initial data.  
However, the process of adding matter to a Kerr black hole and solving 
the initial constraints would also introduce radiation.  Therefore, we 
choose the simpler initial data set provided by Bowen and 
York~\cite{Bowen80}.

Following~\cite{Brandt94b}, we choose
\begin{eqnarray}
{\hat H}_E = 3J,
\hspace{2cm}
{\hat H}_F=0,
\end{eqnarray}
\noindent where $J=a M$ is the total angular momentum of the 
spacetime.  We set initially $J=20$.  We use a grid of $300$ radial 
zones and $30$ angular zones, with $\eta_{max}=6.5$.  With the same 
shell parameters of the previous section we have an ADM mass of 
$6.87M$ which gives a maximum rotation parameter of $a/M=0.42$.  For 
comparison, we note that without the presence of matter this black 
hole system has an ADM mass of $4.75M$.  Hence, the surrounding matter has 
a significant effect on the black hole spacetime.  However, the resulting 
black hole is not necessarily highly distorted, as it could result in 
a larger black hole with a small perturbation.


We evolve the initial data up to a time of $100M$.  By this time the 
final aspect of the lapse, metric function $A$, total velocity and 
density look very similar to the non-rotating run displayed in 
Fig.~\ref{shell1}.  However, at much earlier times of the evolution, 
all variables present a more clear angular dependence as a consequence 
of the rotation.  This is more pronounced as $J$ increases.  As an 
example we plot in Fig.~\ref{rotshell0} the metric component $A$ at 
$t=25M_{ADM}$.  

The complexity of our system is now clearly greater than in 
previous models as
we have additional quantities to evolve, e.g., the metric
function $F$ or the $\phi$ component of the shift. The $\phi$
component of the proper 4-velocity of the fluid is now different
from zero as well. However, the corresponding 3-velocity component,
as defined by 
\begin{eqnarray}
v^i=\frac{u^i}{\alpha u^t} +
\frac{\beta^i}{\alpha},
\end{eqnarray}
\noindent should be zero, after being corrected by the shift.  We find 
this to be the case in our numerical integration.

We also use this 
run to test the ability of the code to conserve angular momentum as
measured by Eq.~\ref{angularmom}.  We 
find that the initial angular momentum of the spacetime is accurately 
maintained to within $5\%$ of the initial value, $J=20$, at all points 
on the grid after $t=100 M$ and converges away to second order.
This behavior is indeed what one expects 
as there is no physical viscosity in the dust shell that could 
transfer the angular momentum.  In fact, as stressed previously, the 
matter in this case is not carrying angular momentum, which can be 
seen from our initial data choice $S_{\phi} =0$.
The matter fields actually are measured by to the so-called 
Zero-Angular-Momentum-Observers (ZAMO's), which although rotating, do 
not carry angular momentum.

We also compare the effects of the rotation and matter fields on the 
shape of the horizon.  This is a further analysis technique that 
proved very useful in vacuum spacetime, which has also seen some use 
in matter spacetimes as well. Smarr~\cite{Smarr73b}
showed that for vacuum Kerr spacetimes, the horizon 
had an oblate shape parameterized by the spin parameter 
$a$. For larger $a$, the horizon becomes more oblate, 
as one might expect from naive considerations of spinning 
objects bulging at the equator.  Previously, we found that for dynamic 
rotating black holes, the horizon oscillates about this oblate shape, 
settling down to its equilibrium value expected for a Kerr black hole 
of the appropriate mass and angular momentum.  In fact, simply by 
measuring the horizon shape, one could determine its mass, angular 
momentum, and oscillation 
frequency~\cite{Brandt94a,Brandt94b,Brandt94c}.  We now apply this 
technique to black holes surrounded by accreting matter.

For this purpose we plot in 
Fig.~\ref{rotshell0_1} the ratio of the polar to the equatorial 
horizon circumference ($C_r=C_p/C_e$) for a sample of four runs: a 
vacuum run with $J=10$, a run with low mass density and $J=10$, a run 
with high mass density and $J=10$, and a run with high mass density 
and $J=0$.  For each plot we also include a straight horizontal line 
corresponding to the value of $C_r$ for a Kerr black hole with the 
same value of $a/M_{ADM}$.  The low density matter distribution is 
given by Eq.~(\ref{shell}) with parameters $\rho_b=10^{-2}$, 
$\kappa=1$, $\eta_0=2$ and $\rho_{max}=1$.  Correspondingly, the high 
density matter distribution has the same values of $\kappa$ and 
$\eta_0$ but the parameters $\rho_b$ and $\rho_{max}$ are ten times 
bigger.

The point to notice in Fig.~\ref{rotshell0_1} is that only the vacuum 
case settles down to the expected value of $C_r$ for the Kerr 
spacetime given its angular momentum and ADM mass.  The others settle 
down to something slightly different.  In the low mass density case it 
is a little less spherical whereas in the high mass density case it is 
something a little more spherical.  We expect that if all matter had 
fallen in the black hole, the standard Kerr result would be obtained.  
Clearly, the spacetime must be settling down to some quasi-stationary 
solution that corresponds to Kerr surrounded by matter.  This is a 
very interesting point that should be explored further in future work.  
The effect of matter around a black hole on its geometry and 
oscillation structure has not received much attention, yet it could 
have important astrophysical consequences.  As gravitational wave 
detectors begin to see waves from black holes, a particularly 
intriguing possibility is that they may carry information about not 
only the black holes themselves, but also about the astrophysical 
environment surrounding them~\cite{Leung97}.


The rotational implosion of the shell induces the emission of odd-parity
gravitational waves in addition to even-parity modes. We plot in 
Figs.~\ref{rotshell1}, \ref{rotshell2} and \ref{rotshell3} the
$\ell=2, 3$ and 5 modes of the emitted radiation with a fit against
the first two harmonics of the corresponding quasinormal modes.  
Again we note that these fits are made to the known 
quasinormal modes of {\em vacuum} black holes, although not all matter 
has crossed the horizon by this time.  Further work should be done to 
study the effect of a ``dirty'' environment on the mode structure of 
black holes~\cite{Leung97}.


\section{Conclusions}
\label{conclusions}

We have presented a new numerical code to study the evolution of 
matter in black hole axisymmetric spacetimes in general relativity.  
Despite the well known ``axis instability" of general relativistic 
axisymmetric codes we are able to evolve, during a reasonable amount 
of time into the future, different initial matter configurations.  The 
two building blocks of the code, spacetime and hydrodynamics, are 
fully coupled through the source terms (and fluxes) of both systems.  
The extreme dynamical range of variation of the metric quantities, as 
shown typically in the peak of the radial function $A$, the grid 
stretching or the formation of singularities, make the hydrodynamical 
integration quite a difficult enterprise.  Within a dynamical 
spacetime framework, the computation becomes much more challenging 
than in the unrealistic case of an static background gravitational 
field, but it allows one to study the complete problem, including the 
effects of the matter on the black hole evolution and its corresponding 
structure and emission of gravitational waves. As this paper focuses 
on the development and testing of such a code and associated analysis 
tools to study the resulting physics, we defer detailed application to 
astrophysical scenarios to future investigations.

The integration scheme used in the code is based on finite 
differencing the partial differential equations.  For the hydrodynamic 
equations we have used an advanced {\it high-resolution 
shock-capturing} scheme built on approximate Riemann solvers.  The 
integration of the ADM field equations was done with an standard 
explicit second order Runge-Kutta scheme with centered differencing.  
We have presented convergence tests of the code as well as a 
sufficient set of astrophysical applications.  These include the 
spherical accretion of matter onto a black hole, the implosion of dust 
shells and the evolution of matter in a rotating black hole spacetime.  
We have also computed the waveforms induced by the presence of the 
matter in some of the aforementioned simulations.

Because dynamic black holes accreting matter have not been studied 
previously, we developed a new series of testbeds appropriate for this 
problem and applied them to our code.  (a)  Building on the standard Bondi 
accretion on static black hole metric, which is an analytic solution, 
we showed how one can compare the numerical solution obtained by 
on a dynamically sliced background.  (b)  We also showed how one can 
computed radiation waveforms from the fully coupled matter-black 
hole system, which are emitted as the accretion induces oscillations 
in the black hole spacetime.  (c)  We applied a set of analysis tools 
developed to study the properties of black hole horizons in vacuum 
spacetimes to the accretion problem, and found them to be useful in 
studying the energy accounting of the entire black hole plus matter plus
radiation system.  We also studied the geometry of the black hole 
horizon as it is distorted by the presence of matter falling in.

One particularly interesting point which emerges from these studies is the 
possibility that the matter surrounding the black hole perturbs it 
in a measurable way.  The geometry of the hole is seen to 
be changed by the presence of matter, and it is possible that the 
radiation structure that one hopes to measure ultimately may be 
affected as well\cite{Leung97}.  This will need much more study in the 
future; a fully coupled hydrodynamics
and spacetime code like the one developed 
here can address this problem in its full non-linearity.

In subsequent work we plan to extend the results presented here 
performing detailed comparative and parametric studies of the 
different scenarios just outlined in the present investigation, 
including axisymmetric (non-spherical)
accretion onto black holes and its effect on the structure of the black
hole geometry and on the gravitational radiation emitted,
the head-on collision of 
stars with black holes and a detailed comparison with recently 
developed perturbative treatments of matter flows around black 
holes~\cite{Papadopoulos98c}. We also plan to use these results and this code 
as a testbed for future computations with the three-dimensional 
coupled code, called Cactus, we are currently 
developing~\cite{Bona98b,Font98b}.

\section{Acknowledgments}
We have benefitted from numerous conversations with our colleagues at 
AEI and elsewhere, particularly Philippos Papadopoulos and Peter Anninos.
This work was supported by AEI.  J.A.F acknowledges financial support from the 
TMR program of the European Union (contract number ERBFMBICT971902).  

\bibliographystyle{prsty}

\newpage


\begin{table}
\begin{tabular}{|c|c|c|c|}
var & 2-pt conv & var & 3-pt conv \\
\hline
Ham & 1.81 & $\rho$ & 2.04 \\
Mom${}_\eta$ & 1.97 & $S_\eta$ & 1.76 \\
Mom${}_\theta$ & 1.44 & $A$ & 1.83 \\
Mom${}_\phi$ & 2.00 & $B$ & 1.68 \\
tr$K$ & 2.04 & $D$ & 1.89 \\
\end{tabular}
\caption{The result of a convergence test for the problem of an
imploding dust shell with a rotating black hole
with parameters $(\rho_b,\kappa,\rho_{max},\eta_0,n,J)=
(0.01,1,3,2,2,10)$. The initial data is evolved for $6 M_{ADM}$ on grids of
size $100\times10$, $200\times20$, and $400\times40$
in $\eta$ and $\theta$, respectively. Convergence is nearly
$2$ for most quantities. The $\theta$-momentum constraint is
most affected by the axis instability and shows the lowest
convergence. The number given is the average convergence
value along the line $\theta=\pi/4$ for each quantity.}
\label{tab:converge}
\end{table}


\vspace{0.7cm}

\vbox{\begin{figure}
\incpsf{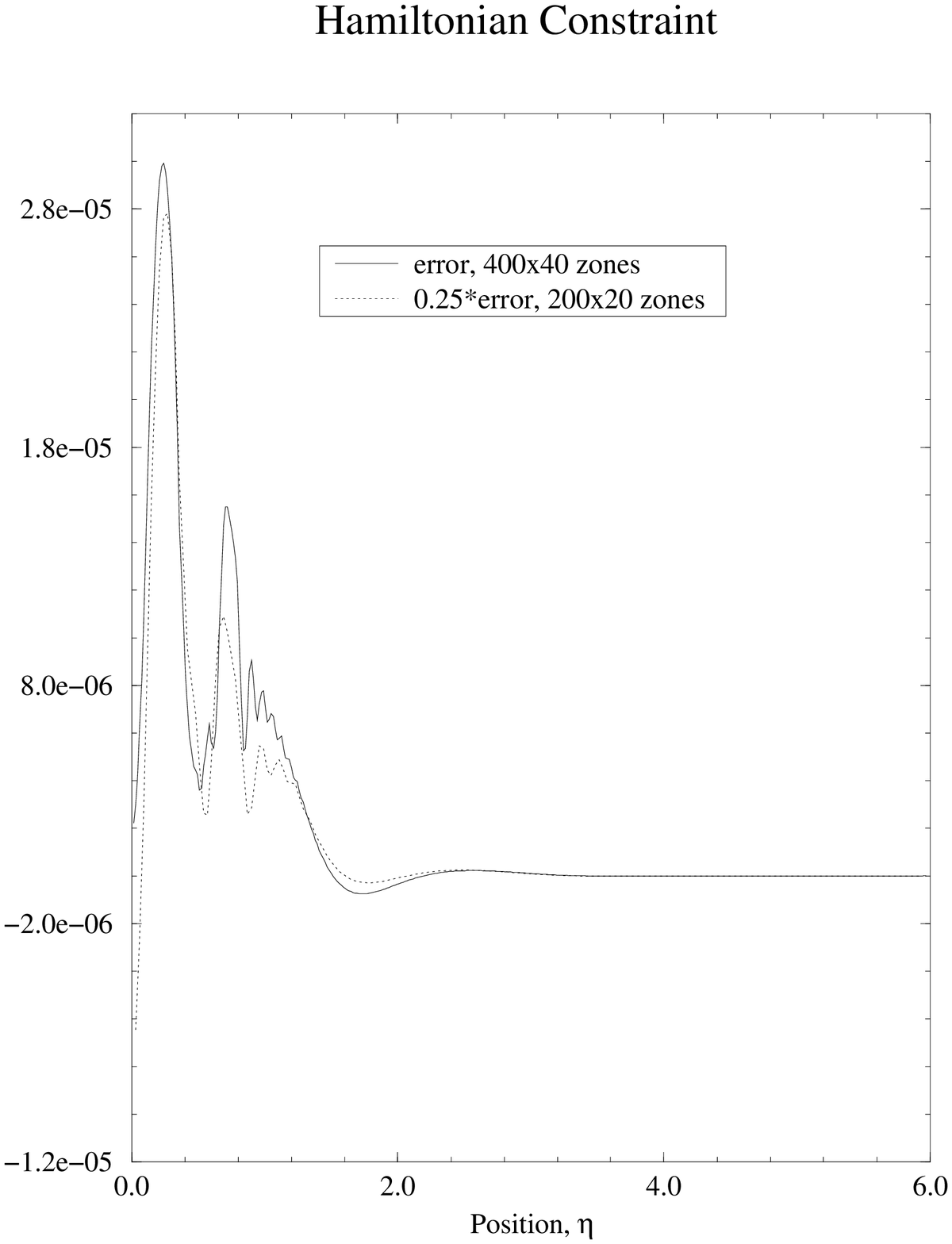}
\caption{Numerical violation of the Hamiltonian 
constraint for the implosion of a dust shell onto a
rotating black hole with $J=10$, as measured along
$\theta=\pi/4$.  We plot the constraint violation at
high resolution and 1/4 the constraint violation at
medium resolution.  For second order convergence these
curves should lie on top of one another.}
\label{fig:ham}
\end{figure}}

\vspace{0.2cm}

\vbox{\begin{figure}
\incpsf{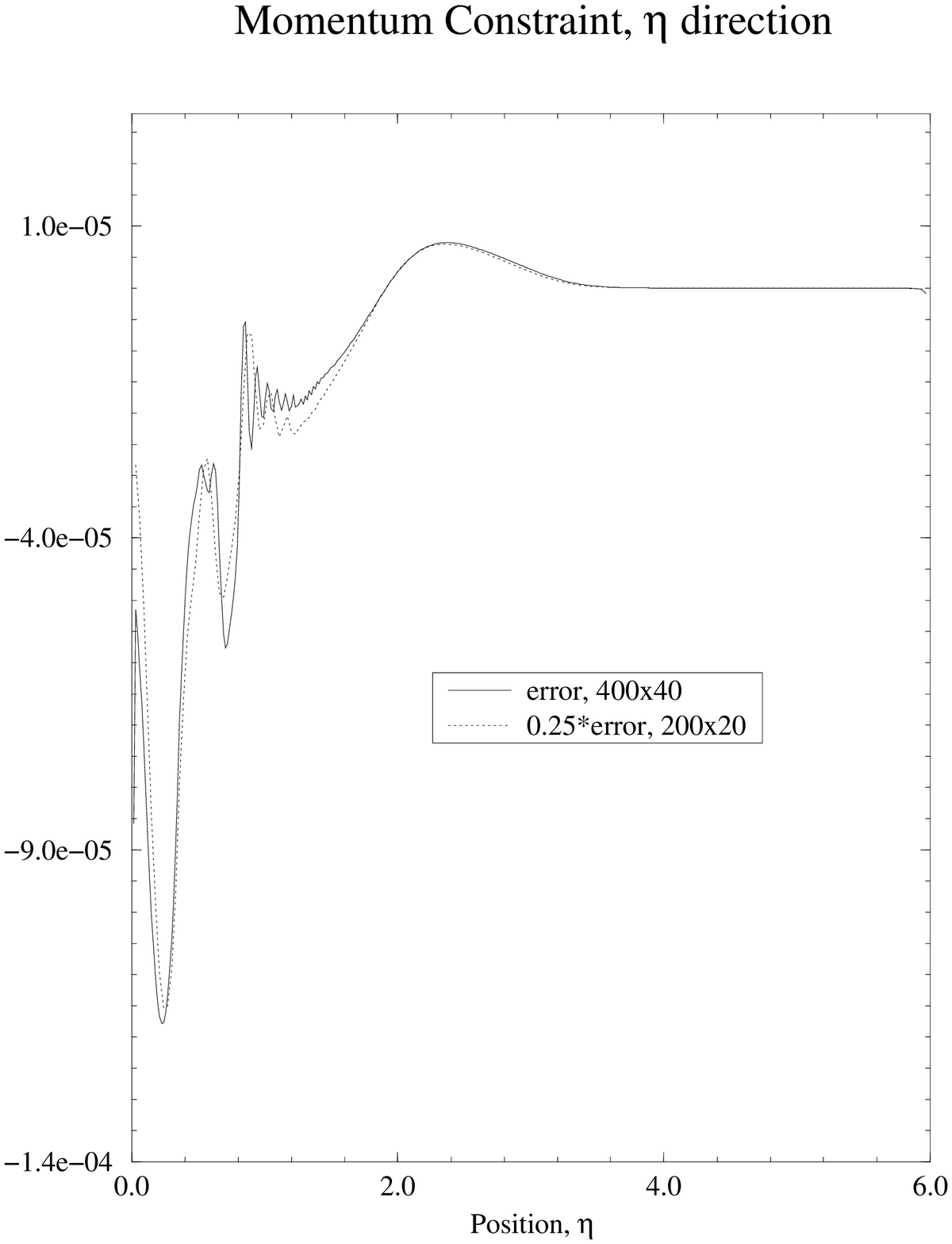}
\caption{Numerical violation of the $\eta$-component
of the momentum constraint for the implosion of a dust shell onto a
rotating black hole with $J=10$, as measured along
$\theta=\pi/4$.  We plot the constraint violation at
high resolution and 1/4 the constraint violation at
medium resolution.  For second order convergence these
curves should lie on top of one another.}
\label{fig:mom1}
\end{figure}}

\vspace{0.2cm}

\vbox{\begin{figure}
\incpsf{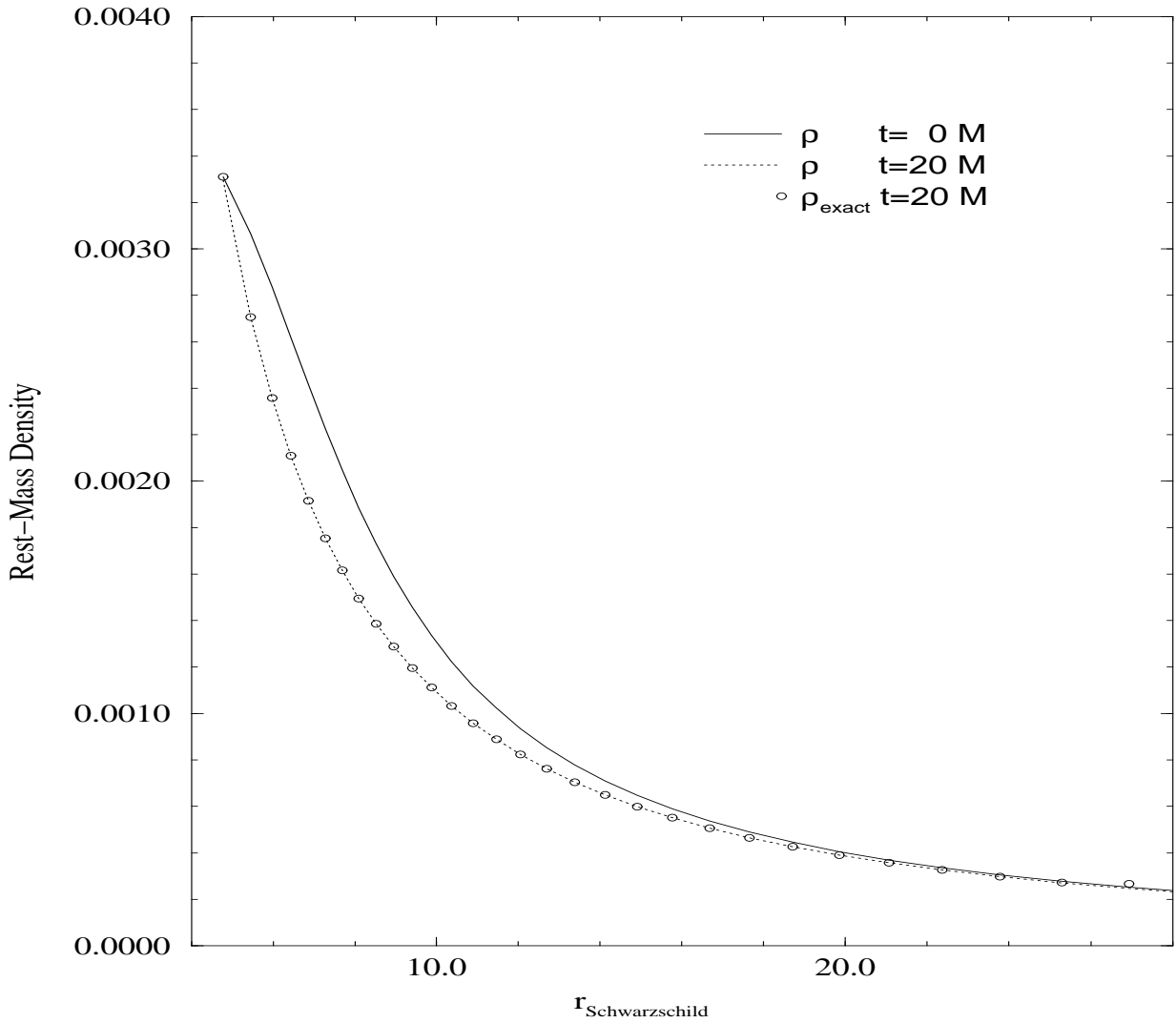}
\caption{This figure shows the rest-mass density for a dynamically
sliced spherical accretion of dust on a Schwarzschild
background.  The solid line represents the initial value -- which
is also the static analytic solution.  The dotted line shows the
density at a later time when the background spacetime is allowed
to evolve with a dynamic slicing condition.  The circles give the
analytic density function but using $\sqrt{g_{\theta\theta}}$ to
calculate the Schwarzschild radius. The functions are identical
to one part in $10^{-5}$. }
\label{BondiComp}
\end{figure}}

\vspace{0.2cm}

\vbox{\begin{figure}
\incpsf{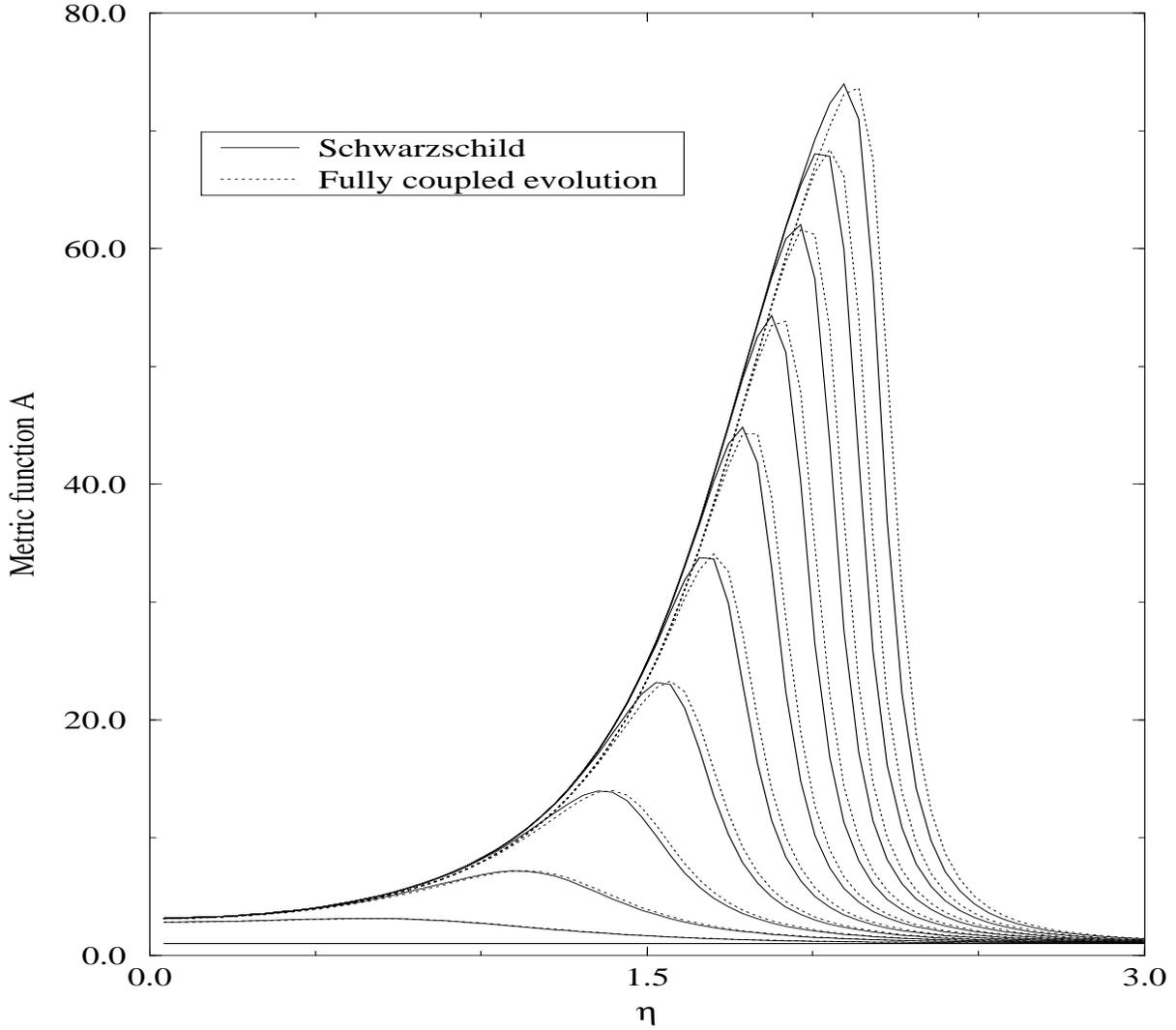}
\caption{Time evolution of the metric component $A$ for two different
{\it couplings} between the hydrodynamics and metric components of the
code. The solid lines indicate Schwarzschild (i.e., matter flowing
on a dynamically sliced background which cannot be affected by the
matter fields) and the dotted
lines correspond to a {\it ful coupling} 
(i.e., fully self-gravitating matter that is allowed to affect
the spacetime geometry). The
last curve corresponds to $t=50M_{ADM}$. The initial matter density is
$\hat{\rho}=10^{-2}$. Almost no differences are found between the two 
evolutions for such low density matter flows.}
\label{dynslice1}
\end{figure}}

\vspace{0.2cm}


\vbox{\begin{figure}
\incpsf{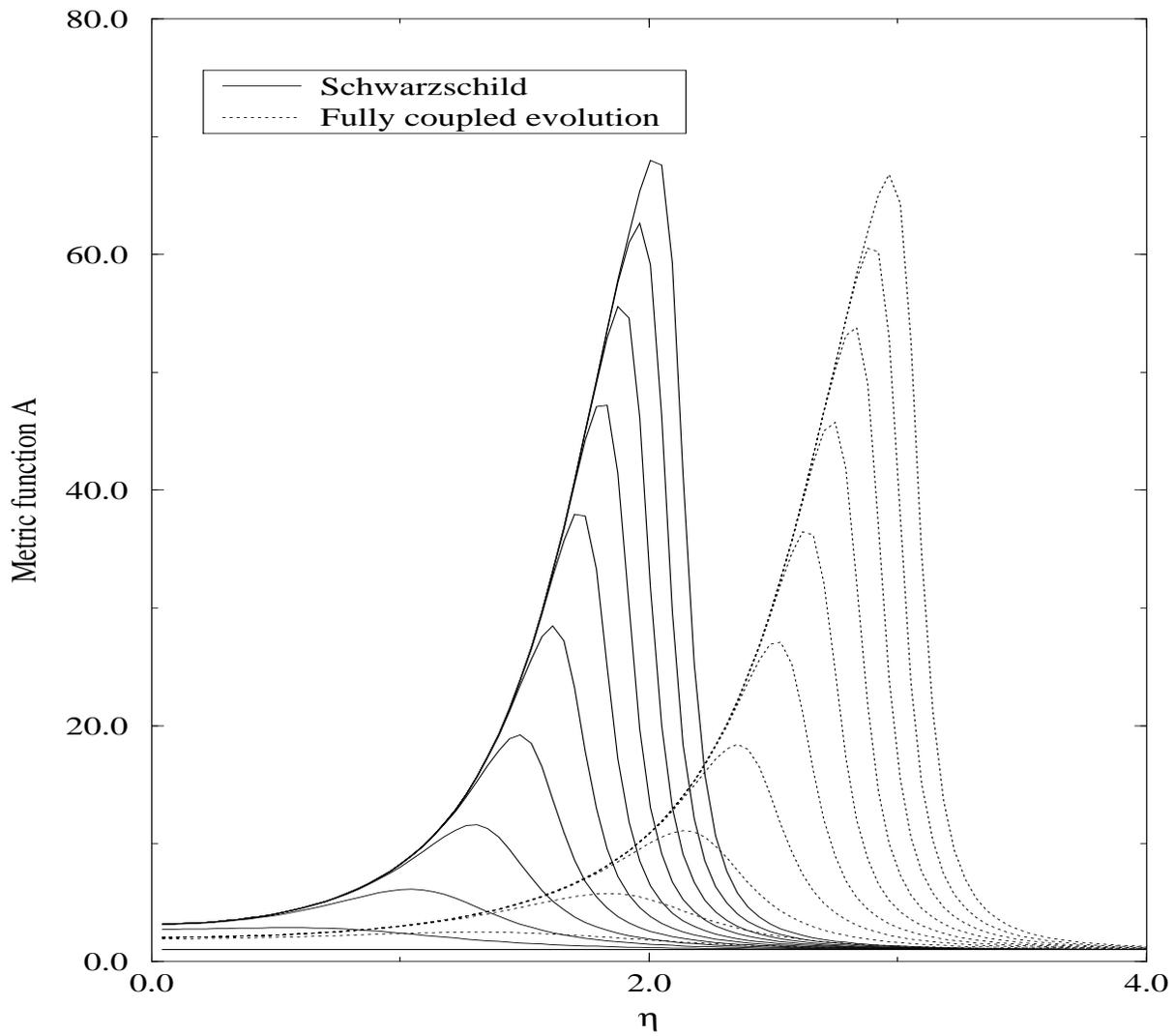}
\caption{Same as Fig.~\ref{dynslice1} but for an initial 
uniform density distribution $\hat{\rho}=1$. Now, the evolution of
the spacetime is noticeably different if the metric reacts or not
to the presence of the matter.}
\label{dynslice2}
\end{figure}}

\vspace{0.2cm}


\vbox{\begin{figure}
\incpsf{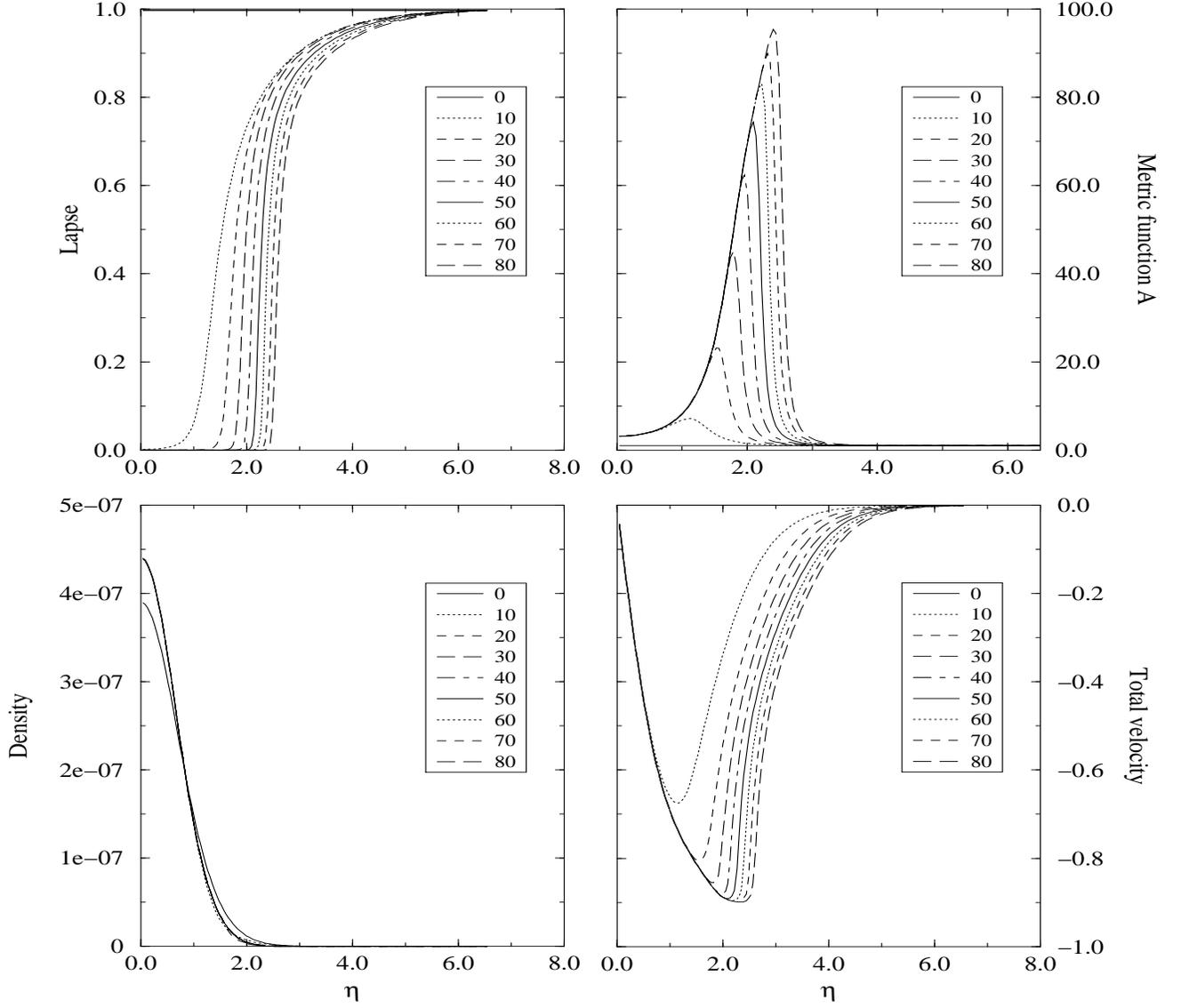}
\caption{Evolution of different metric and hydrodynamical quantities
for the problem of spherical accretion of dust with $\hat{\rho}=10^{-2}$.
The {\it top} panels show the lapse ({\it left}) and the metric
component labelled $A$. The {\it bottom} plots show the rest-mass
density ({\it left}) and the total velocity of the dust particles.}
\label{sphacc1}
\end{figure}}

\vspace{0.2cm}


\vbox{\begin{figure}
\incpsf{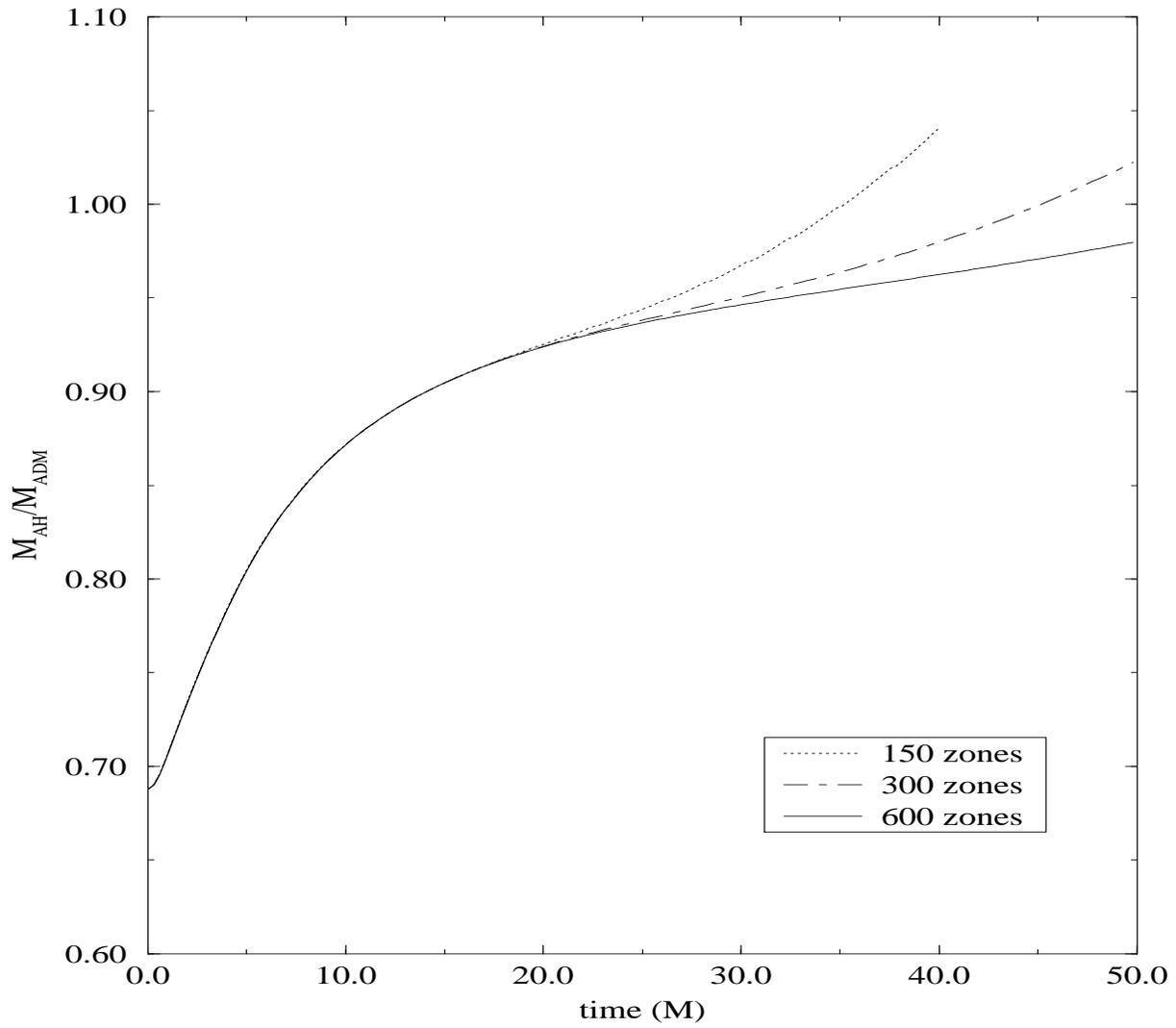}
\caption{Time evolution of the apparent horizon mass for the spherical
accretion of dust problem. Each curve corresponds to a different radial
resolution as indicated in the plot legend.}
\label{sphacc2}
\end{figure}}

\vspace{0.2cm}


\vbox{\begin{figure}
\incpsf{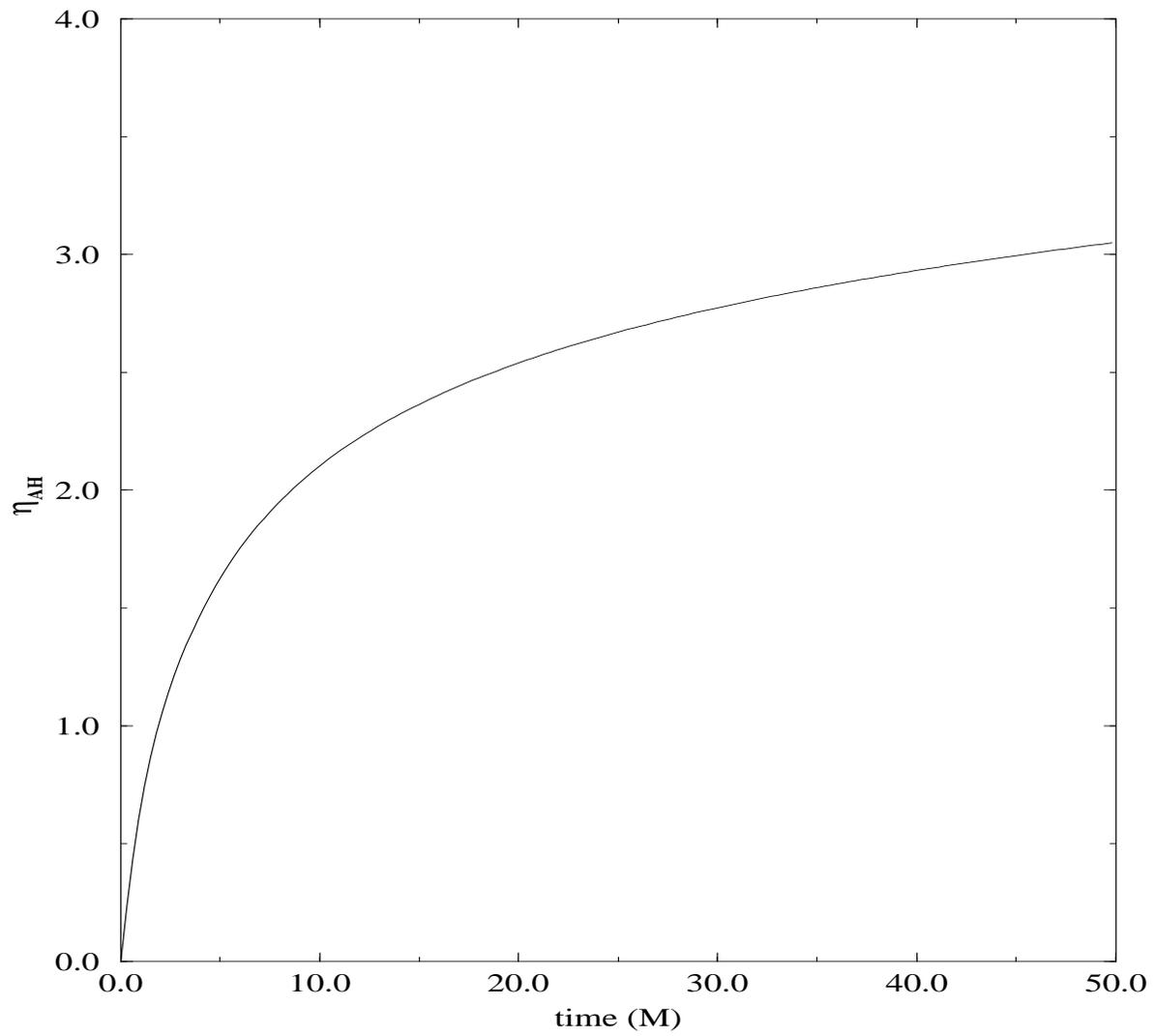}
\caption{Apparent horizon location as a function of time for the
spherical dust accretion problem.}
\label{sphacc2_2}
\end{figure}}

\vspace{0.2cm}


\vbox{\begin{figure}
\incpsf{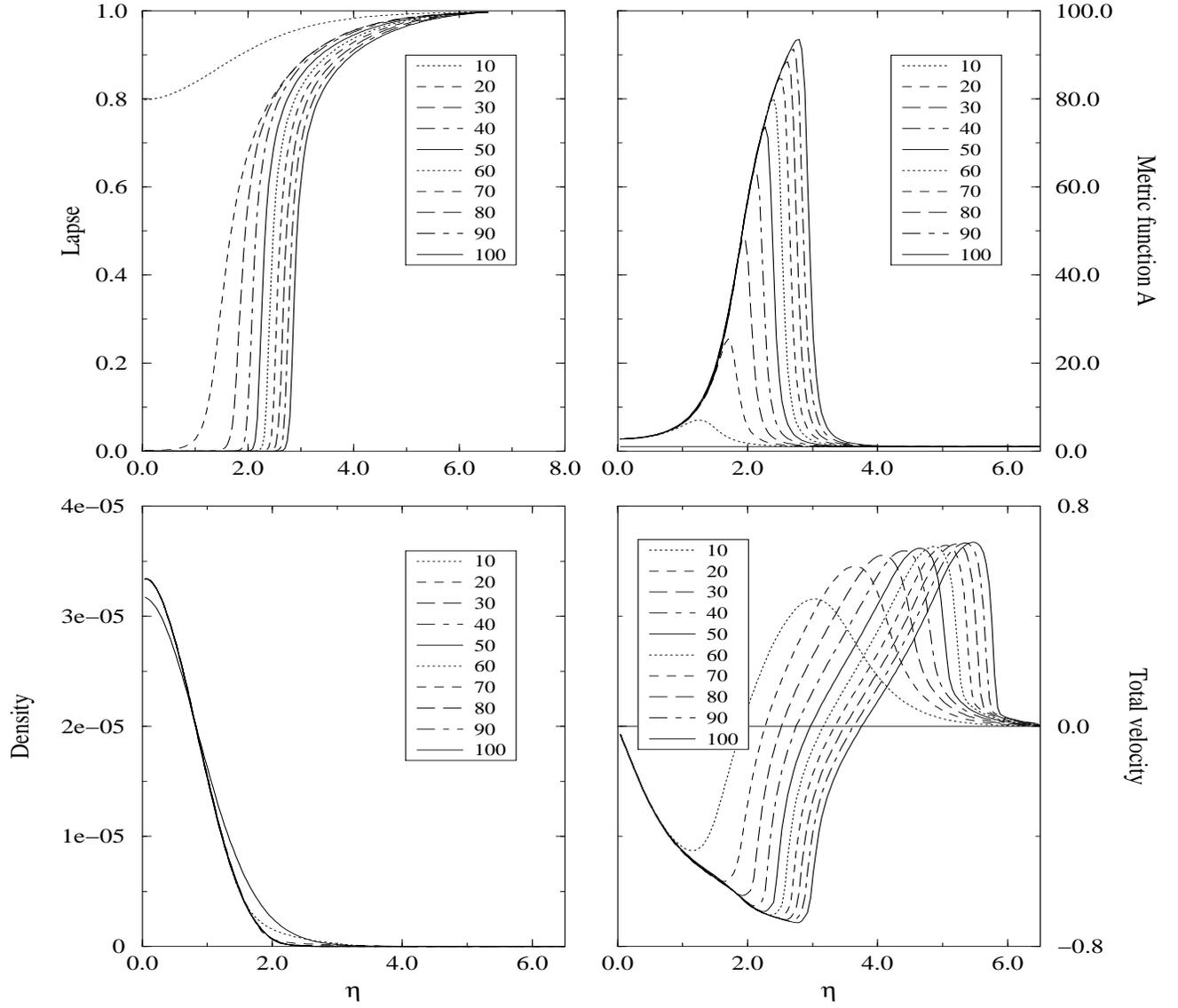}
\caption{Evolution of different metric and hydrodynamical quantities
for the problem of spherical accretion of a perfect fluid.
The {\it top} panels show the lapse ({\it left}) and the metric
component labelled $A$. The {\it bottom} plots show the rest-mass
density ({\it left}) and the total velocity of the fluid.}
\label{sphacc3}
\end{figure}}

\vspace{0.2cm}


\vbox{\begin{figure}
\incpsfw{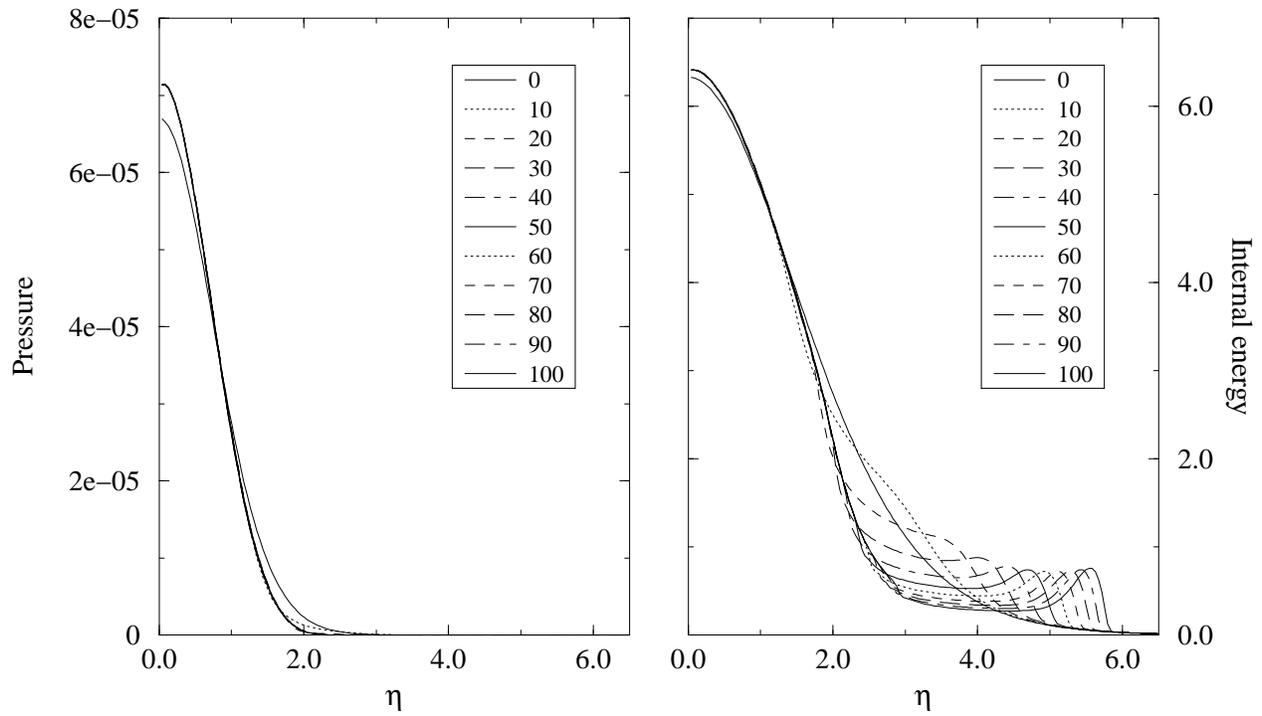}
\caption{Evolution of the pressure ({\it left}) and internal
energy density for the problem of spherical accretion of a
perfect fluid with $\gamma=4/3$.}
\label{sphacc4}
\end{figure}}

\vspace{0.2cm}


\vbox{\begin{figure}
\incpsf{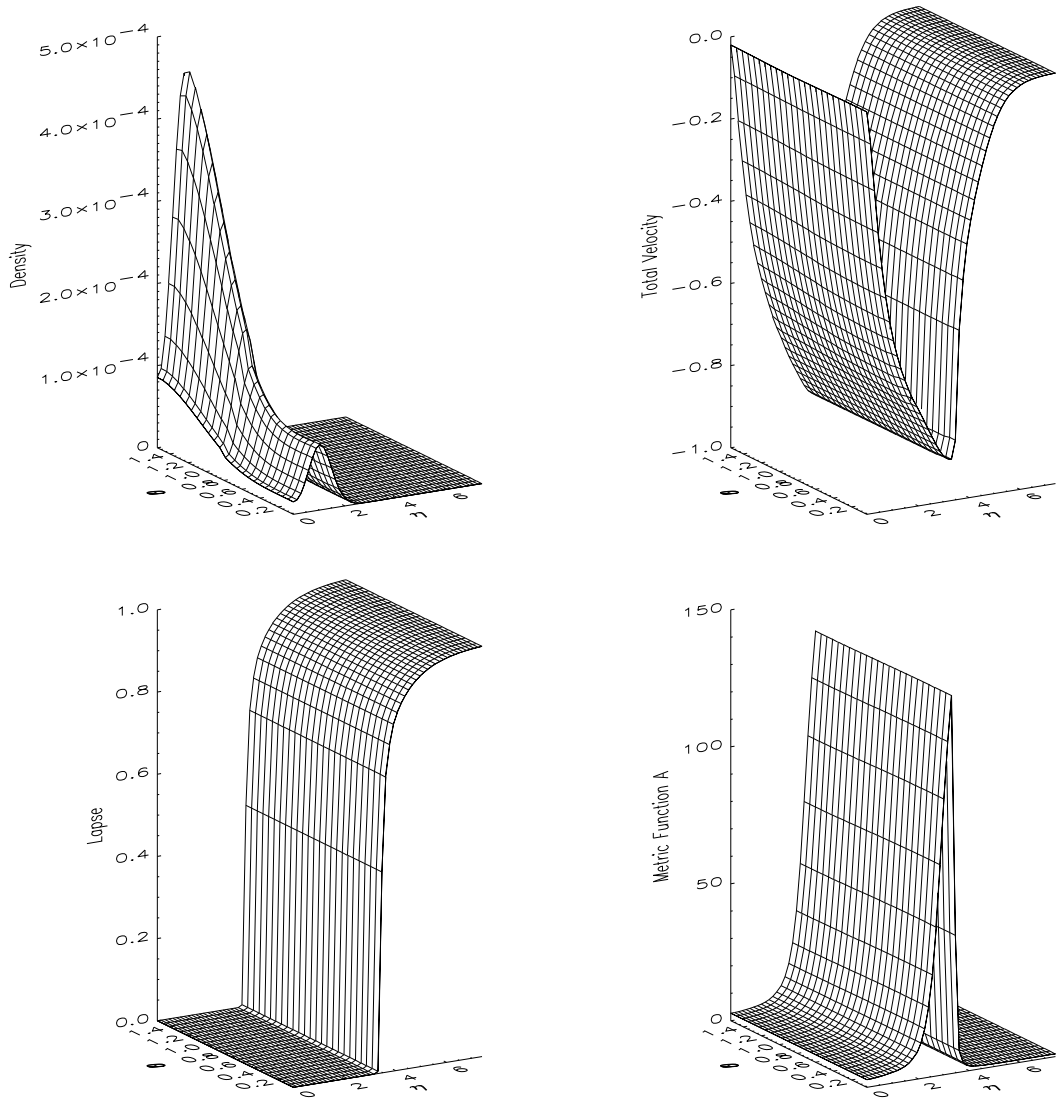}
\caption{Different metric and hydrodynamic quantities for the
impact of an imploding shell of dust with the black hole. The
solution is plotted at a final time of $100M$. From {\it top-left}
to {\it bottom-right} we show the density, total velocity, lapse
and metric function $A$.}
\label{shell1}
\end{figure}}

\vspace{0.2cm}


\vbox{\begin{figure}
\incpsf{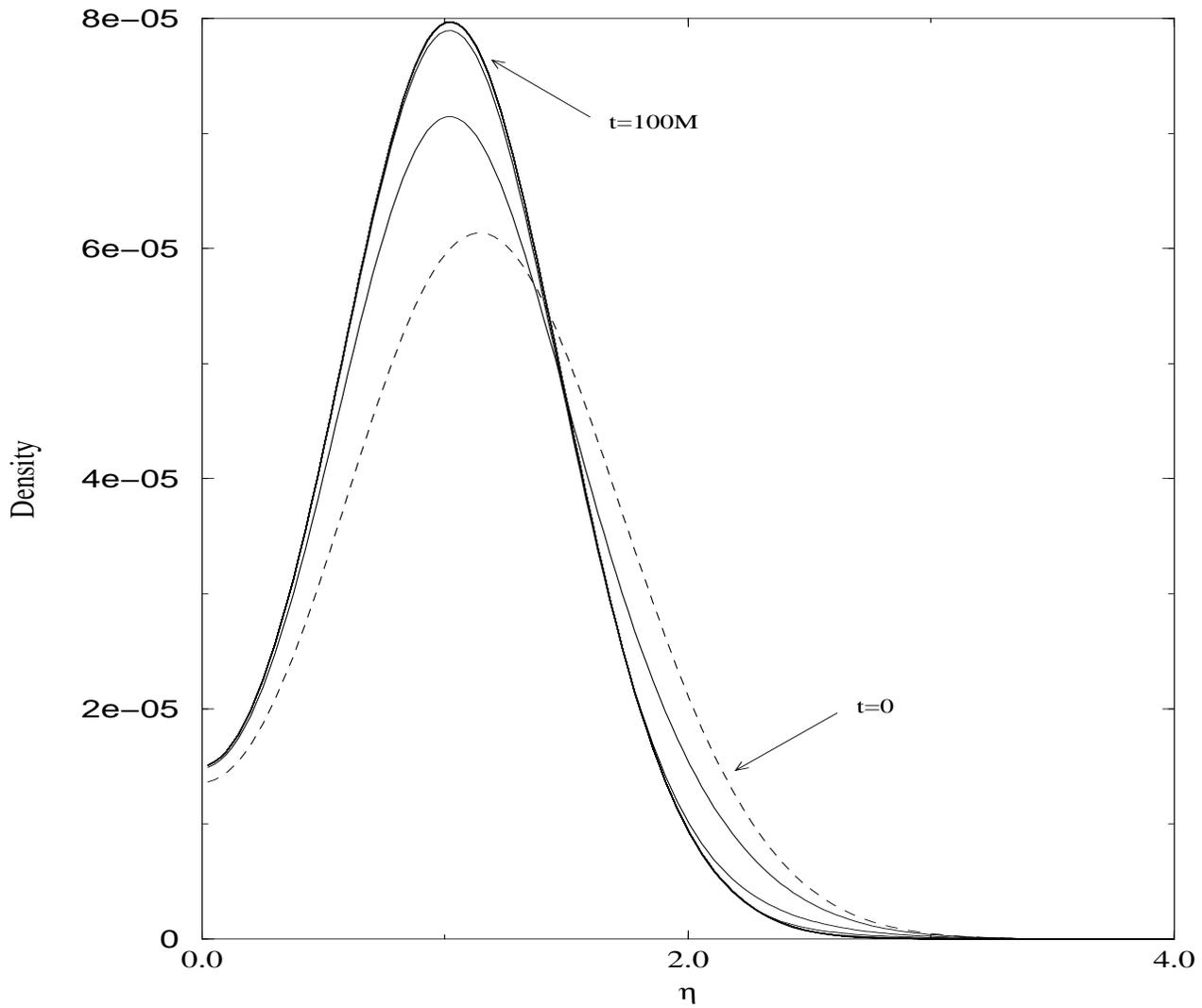}
\caption{Radial profiles of the density at different times of the
evolution of an imploding shell. They correspond to an arbitrary
constant value of the angular coordinate. Notice the 
collapse of the shell in progressive times. The thick solid line
shows the initial profile.}
\label{shell2}
\end{figure}}

\vspace{0.2cm}


\vbox{\begin{figure}
\incpsf{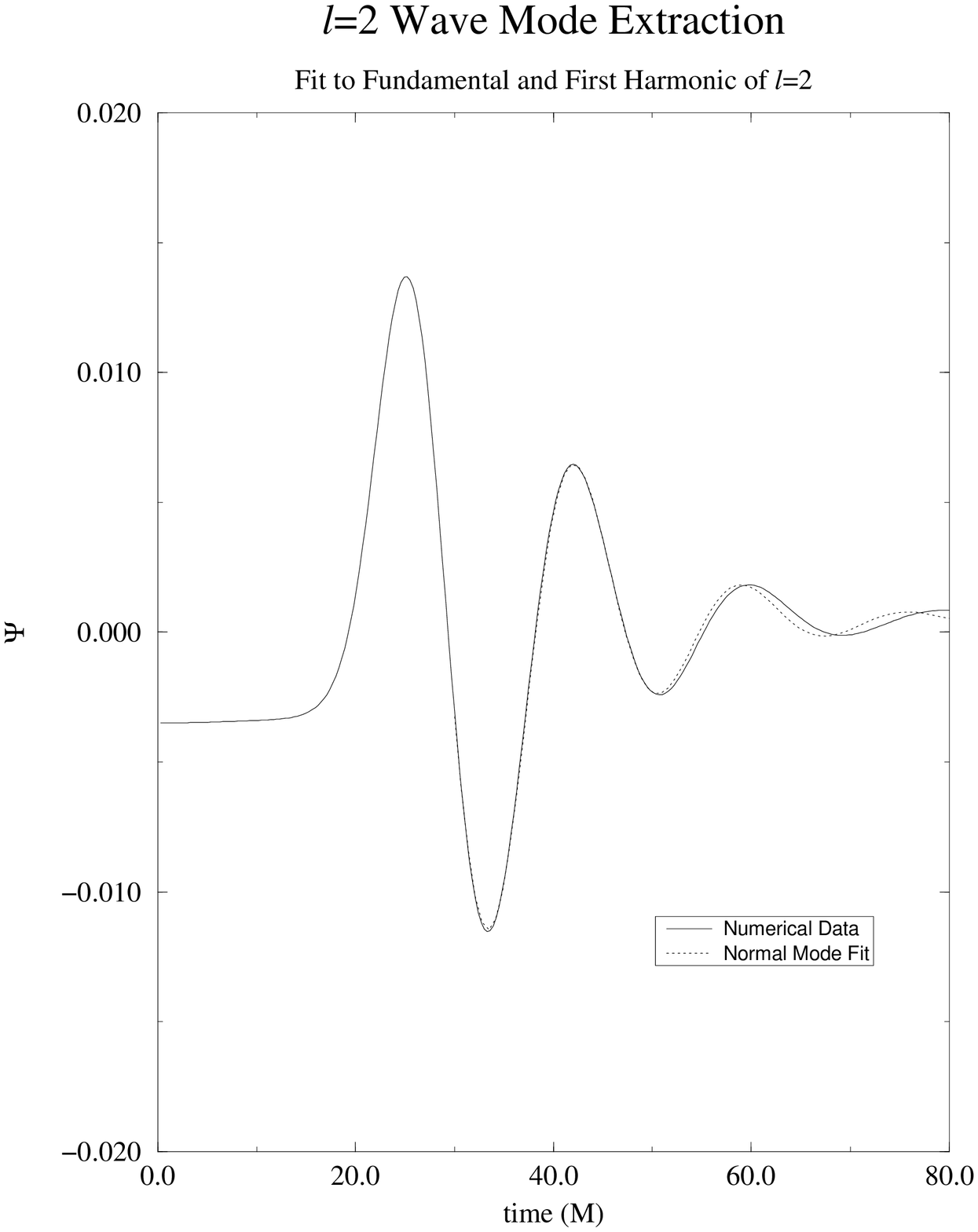}
\caption{This figure shows the numerically extracted $\ell=2$
waveform (solid line) and the least squares fit to the two lowest
$\ell=2$ quasinormal mode for the imploding shell problem.}
\label{shell3}
\end{figure}}

\vspace{0.2cm}


\vbox{\begin{figure}
\incpsf{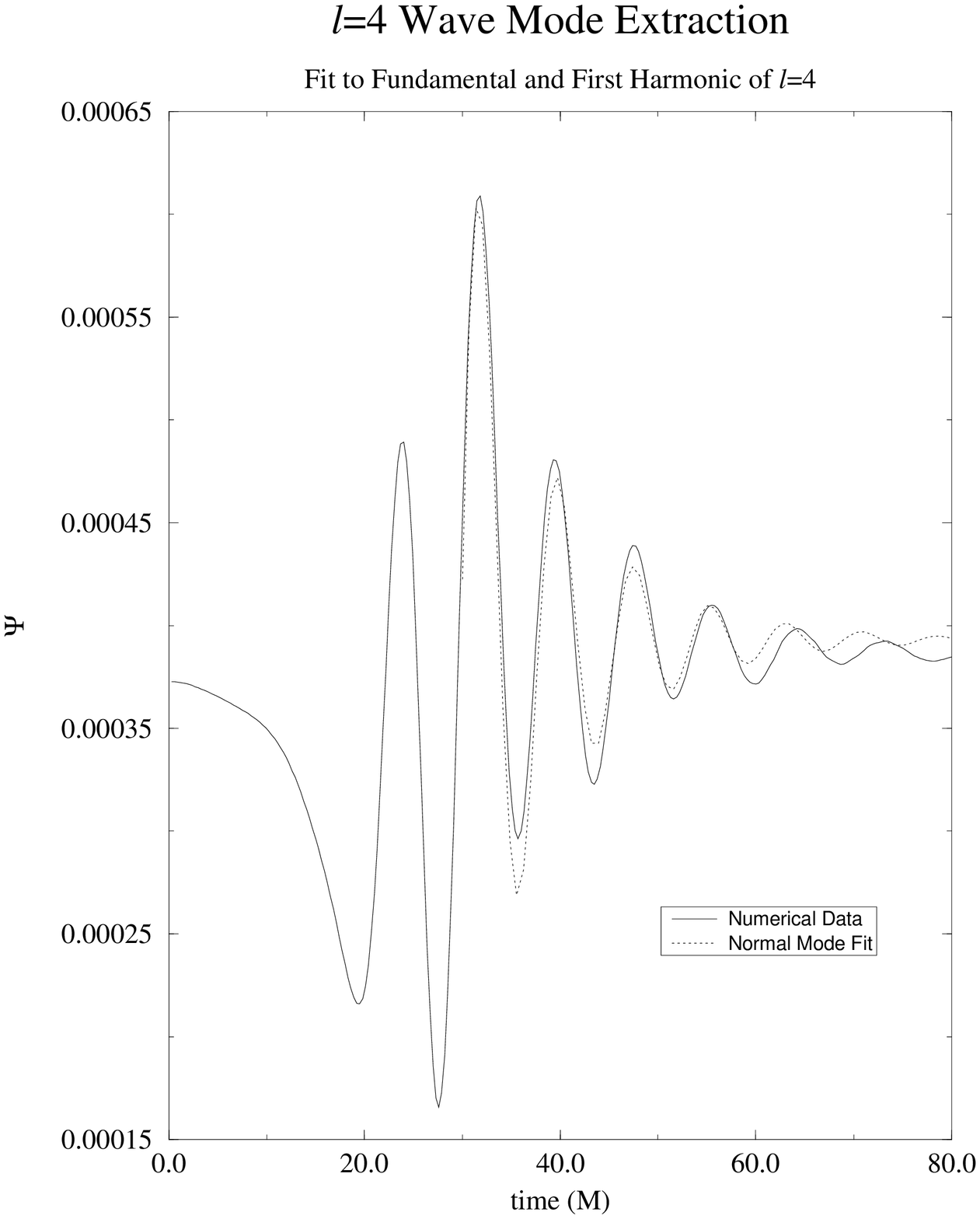}
\caption{This figure shows the numerically extracted $\ell=4$
waveform (solid line) and the least squares fit to the two lowest
$\ell=4$ quasinormal mode for the imploding shell problem.}
\label{shell4}
\end{figure}}

\vspace{0.2cm}


\vbox{\begin{figure}
\incpsf{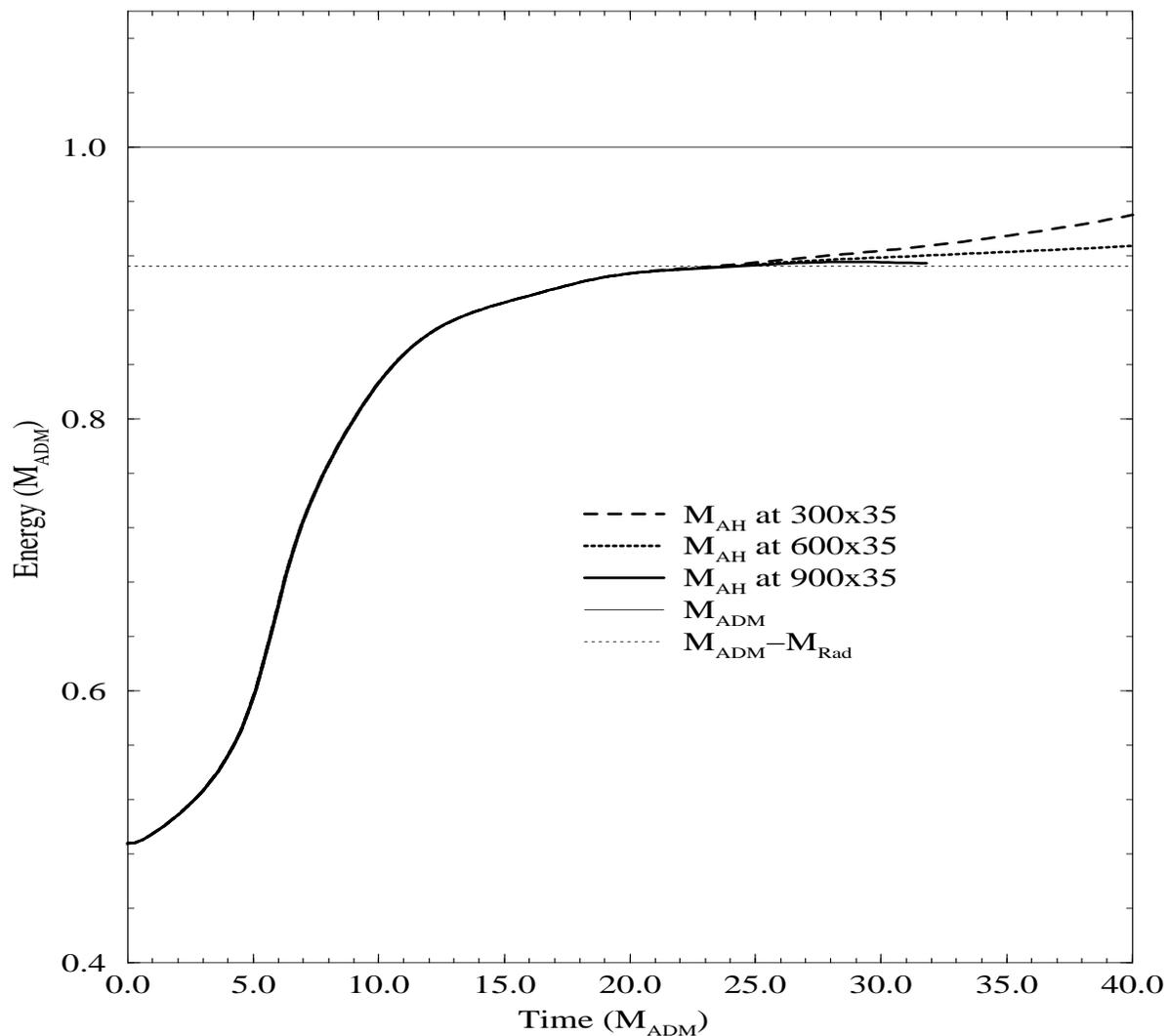}
\caption{This figure shows the energy accounting for a simulation
with a Brill wave and matter fields at a variety of
resolutions.  The Brill wave amplitude is $0.5$, its width is
$1.0$ and its location is $\eta_0=2.5$.  The matter distribution
is located at $\eta_0=2.0$ and has an amplitude of $3.0$. 
The background conformal density is $10^{-4}$.}
\label{account}
\end{figure}}

\vspace{0.2cm}


\vbox{\begin{figure}
\incpsf{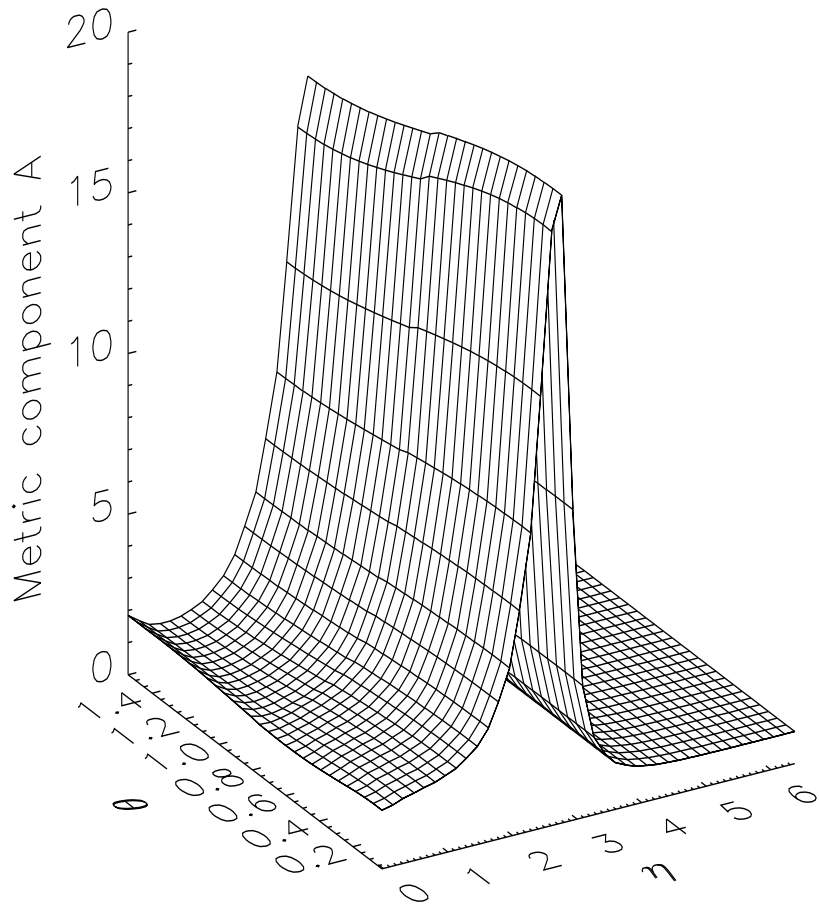}
\caption{Metric component $A$ at $t=25M_{ADM}$ for a rotating black
hole spacetime with $J=20$. Notice the slightly less grid stretching
near the equator as a consequence of the rotation.}
\label{rotshell0}
\end{figure}}

\vspace{0.2cm}


\vbox{\begin{figure}
\incpsf{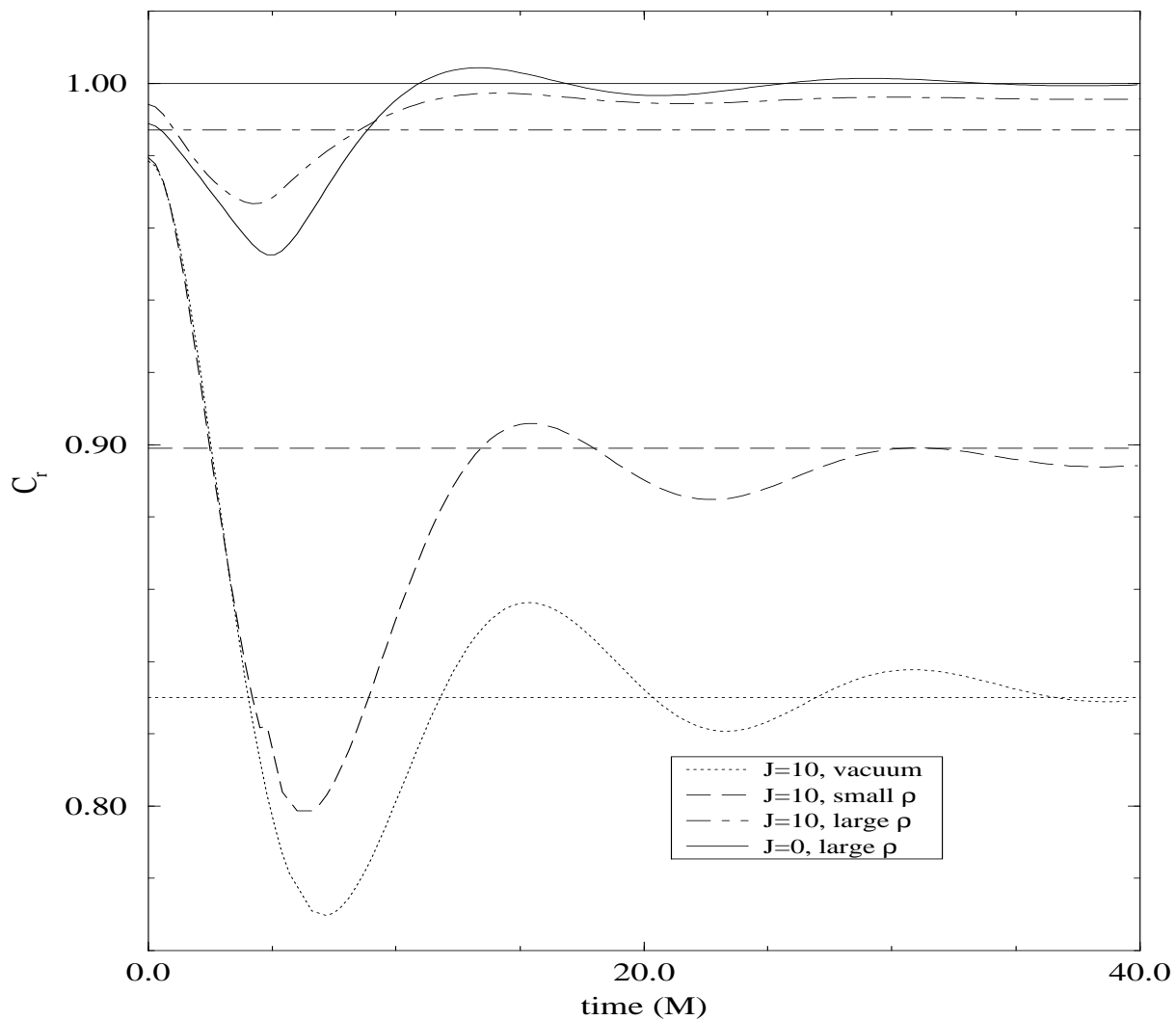}
\caption{The ratio of the polar to equatorial circumference of the
apparent horizon for a sample of rotating spacetimes. In each case,
a horizontal line corresponding to the ratio for a Kerr black hole
with the same ADM mass and angular momentum.  The dotted line
corresponds to a vacuum spacetime while the remaining three curves
correspond to matter evolutions. One can easily see that only in the
vacuum spacetime does the horizon settle down to something with the
same shape as the Kerr black hole.}
\label{rotshell0_1}
\end{figure}}

\vspace{0.2cm}


\vbox{\begin{figure}
\incpsf{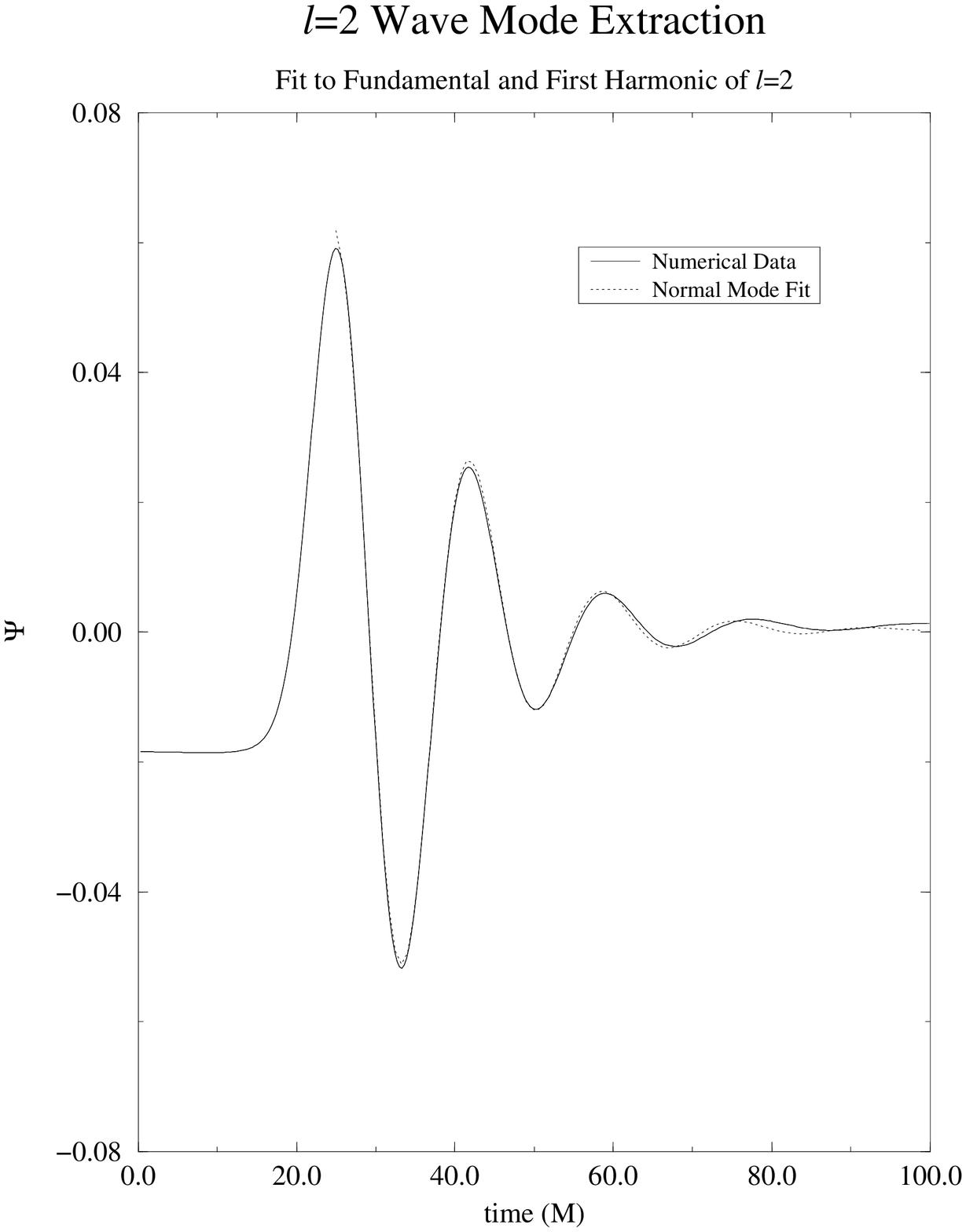}
\caption{This figure shows the numerically extracted $\ell=2$
waveform (solid line) and the least squares fit to the two lowest
$\ell=2$ quasinormal mode for the problem of the implosion of a
dust shell onto a rotating black hole.}
\label{rotshell1}
\end{figure}}

\vspace{0.2cm}


\vbox{\begin{figure}
\incpsf{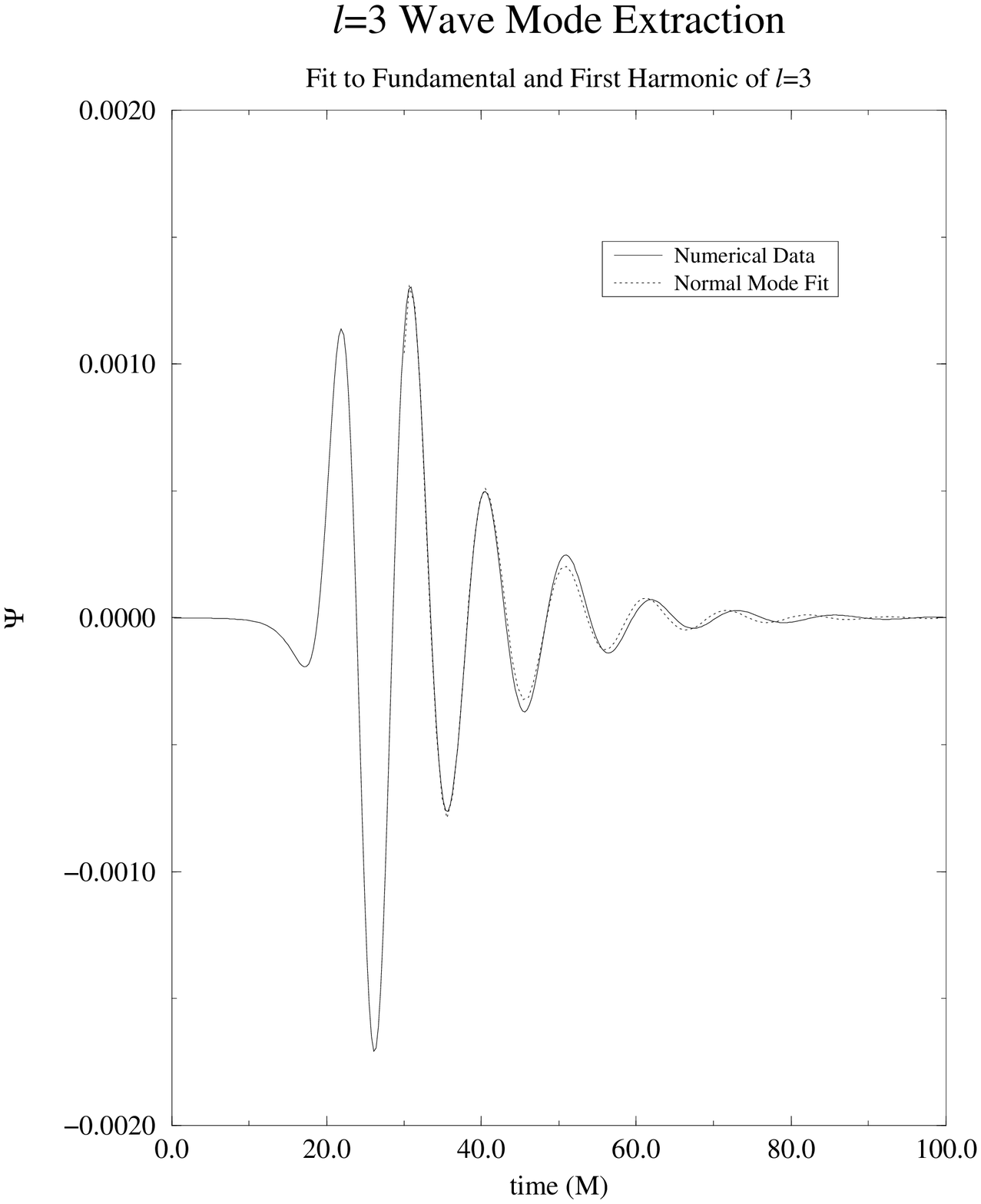}
\caption{This figure shows the numerically extracted $\ell=2$
waveform (solid line) and the least squares fit to the two lowest
$\ell=3$ quasinormal mode for the problem of the implosion of a
dust shell onto a rotating black hole.}
\label{rotshell2}
\end{figure}}

\vspace{0.2cm}


\vbox{\begin{figure}
\incpsf{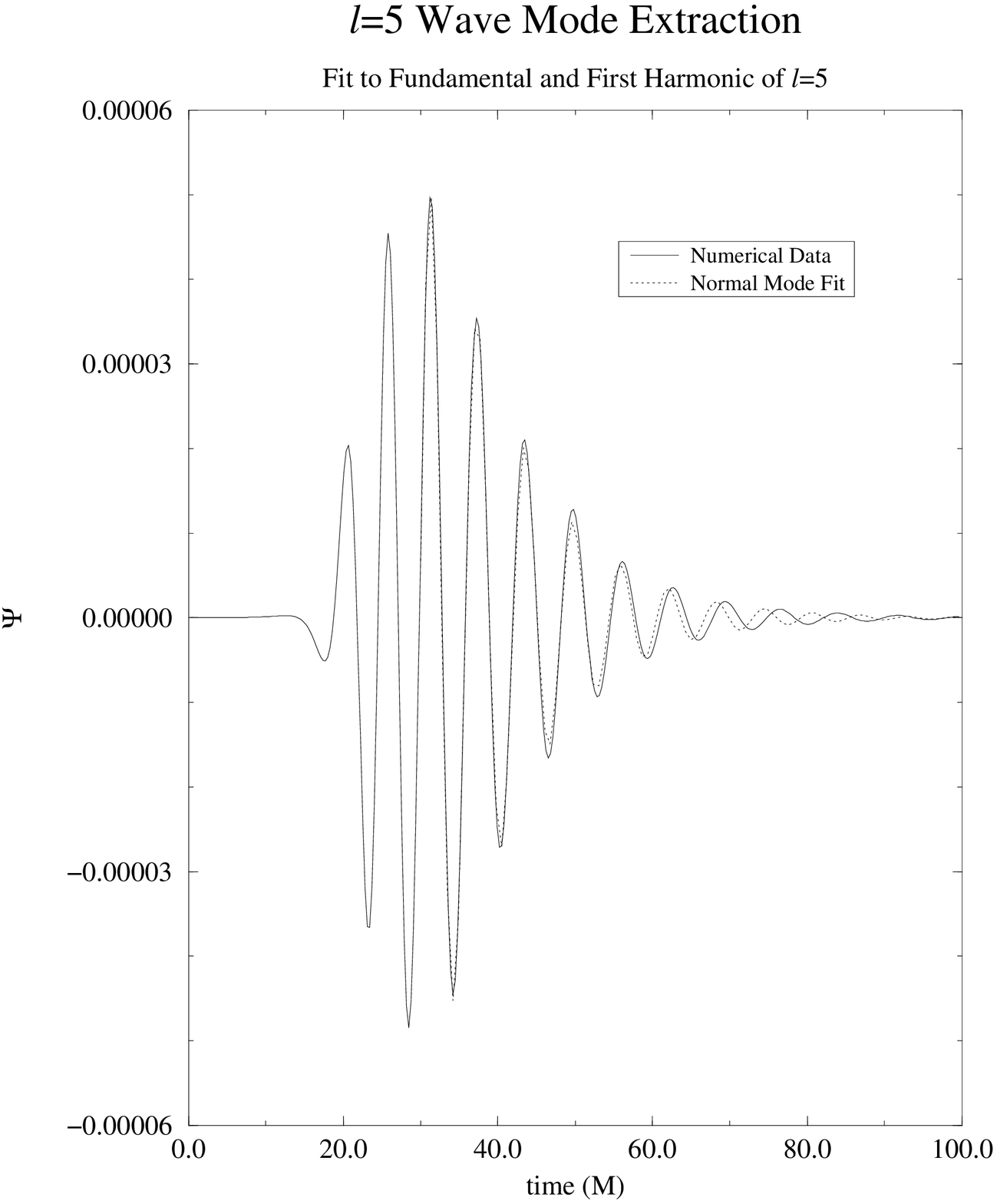}
\caption{This figure shows the numerically extracted $\ell=2$
waveform (solid line) and the least squares fit to the two lowest
$\ell=5$ quasinormal mode for the problem of the implosion of a
dust shell onto a rotating black hole.}
\label{rotshell3}
\end{figure}}

\end{document}